\documentclass[twocolumn,nopacs,prb,preprintnumbers,amsmath,amssymb,aps]{revtex4}
\usepackage{graphicx}% Include figure files
\usepackage{dcolumn}% Align table columns on decimal point
\usepackage{bm}% bold math
\newcommand{\bi}[1]{\mbox{\boldmath$#1$}}

\renewcommand{\theequation}{\arabic{section}.\arabic{equation}}
\def\be{\begin{equation}}
\def\en{\end{equation}} 
\def\p{\partial }  
\def\ve{\varepsilon}
\def\gs{\gtrsim}
\def\ls{\lesssim}

\def\bea{\begin{eqnarray}}
\def\ena{\end{eqnarray}}

\begin{document}
\preprint{APS}

\title{Ion distribution around a charged  rod 
in one and two component solvents: Preferential solvation 
and first order ionization phase transition }% Force line breaks with \\

\author{Ryuichi Okamoto}
\email{okamoto_r@scphys.kyoto-u.ac.jp}
%\author{Akihiko Minami}
\author{Akira Onuki}
\email{onuki@scphys.kyoto-u.ac.jp}
\affiliation{Department of Physics, Kyoto University, Kyoto 606-8502, Japan}

\date{\today}% It is always \today, today,

\begin{abstract}
In one and two component polar solvents, 
we calculate the counterion distribution 
around an ionizable 
 rod  treating the degree of ionization 
$\alpha$ as an annealed variable dependent on its  local environment. 
In the two component  case, 
we take into account the preferential solvation 
of the charged particles and  
 the short-range interaction 
between the rod and the solvent. 
It  follows a composition-dependent mass action law.   
The composition  becomes 
heterogeneous  around a charged  rod 
on a mesoscopic scale, strongly affecting    
the counterion distribution. 
We predict a first order phase transition 
of weak-to-strong ionization       
 for hydrophobic chains. 
This transition  line starts from a 
point on the solvent coexistence curve 
and ends at a critical point. 
The composition heterogeneity is long-ranged near the 
solvent critical point. 
%The degree of ionization tends to zero 
%with increasing the salt density.  
\end{abstract}

\pacs{Valid PACS appear here}% PACS, the Physics and Astronomy
\maketitle
\section{INTRODUCTION}

%%%%%%%%%%%%%%%%%%%%%%%%%%%%%%%%%%%%%%%%%%%%%%%%%
%
%
%%%%%%%%%%%%%%%%%%%%%%%%%%%%%%%%%%%%%%%%%%%%%%%%%

Polyelectrolytes are  extremely complicated 
\cite{Manning,Oosawa,Barrat,Levin,Volk,Holm,Rubinstein,Baigl1,An,Netz,Me}.
Examples include  biological 
polyelectrolytes, like DNA, actin filaments 
or microtubules,  and  synthetic polyelectrolytes. 
In such charged  polymers, 
the  Coulomb repulsion  
among ionized  monomers  
can induce a number of  conformation changes 
of a chain.   The Coulomb  attraction among  
 counterions and an  ionized 
chain  can result in
condensation of counterions at large counterion contents 
(Manning-Oosawa counterion condensation).
In  practice, it is important that 
the phase behavior of polyelectrolytes strongly 
depends on the degree of ionization. 
For example, hydrophobic polymer chains 
become soluble in  water-like solvents  
 with slight ionization.  
In this paper, we further  investigate 
two  complex aspects of polyelectrolytes, 
which have not yet been fully discussed.

First,  the  ionization  (or dissociation) 
process  should be treated as a chemical 
reaction in many polyelectrolytes 
containing weak acidic  monomers 
\cite{Joanny,Holm,Bu1,Bu2,Muth,Burak,Onuki-Okamoto}. 
The degree of ionization $\alpha$ is an annealed 
fluctuating variable governed by the mass action law 
and dependent on 
the local values of  the  
counterion density and the composition (in mixture 
solvents).  All these quantities depend on 
the electric potential  self-consistently. 
Inhomogeneity  of $\alpha$ appears  on  a chain  
\cite{Holm}, but   is   crucial 
in structure formation and 
phase separation \cite{Onuki-Okamoto}. 
Even when  $\alpha$ is nearly homogeneous, 
it is a complex quantity dependent on various 
conditions. It 
can be small in solvents with  low dielectric 
constant and  can increase considerably 
in highly polar  solvents.

Second, complex effects are induced in polyelectrolytes  
when a second fluid component (cosolvent) 
is added to  a water-like solvent. For example, 
precipitation of DNA has been widely observed 
with addition of (less polar) 
alcohol such as 
ethanol to water \cite{Zimm,Bloomfield,Rau,Rauy}.  
Here  the alcohol added  is 
excluded from condensed DNA  
\cite{Rau,Rauy}.  
On the contrary, with 
addition of zwitterionic species (which are more polarizable 
than water), Flock {\it et al.}  \cite{Flock} observed 
 a resolubilization of DNA in the presence of multivalent 
ions (as condensating agents).
 Baigl and Yoshikawa \cite{Baigl} found  
that  zwitterionic  species   
increased stability of coil states  
of a single  DNA  molecule, 
which undergoes  a discontinuous phase transition 
between elongated coil and  compact globule 
depending on the amounts of the second component 
and the multivalent ions (or cationic surfactant 
\cite{Yoshi1}).

To understand these mixture effects, 
relevance of the preferential solvation  
has been pointed out by  experimental 
groups \cite{Bloomfield,Rau,Rauy} 
and  by a theoretical group 
\cite{Andelman1}.   
From our viewpoint,  
particularly  important should be 
the ion-dipole interaction among charged particles 
  and polar molecules \cite{Is}, 
 which gives rise to  the solvation 
(hydration) shell composed of several   solvent molecules 
(those of the  more polar component 
in  a mixture solvent) around each charged particle. 
The resultant  solvation 
chemical potentials  of ions   
 typically much exceed  the thermal energy $k_B T$ 
(per ion) and   strongly  
depend  on the composition in  binary mixtures. 
It is decreased (increased) for hydrophilic (hydrophobic) 
ions with increasing the water composition 
in a mixture of water+  less polar component. 
Furthermore, the degree of ionization of polymers 
can strongly depend on the composition.  
Recently, including such solvation interactions,  
several theoretical groups  have begun to investigate  
the ion effects  in electrolytes 
with mixture solvents 
\cite{Onuki-Kitamura,OnukiPRE,OnukiJCP,Tsori,Roij,Andelman1}, 
polyelectrolytes\cite{Onuki-Okamoto}, 
and ionic surfactants at  oil-water interfaces \cite{OnukiEPL}.

In this work, we will 
demonstrate emergence of mesoscopically  
heterogeneous composition variations 
 around a charged rod, 
which stem from the preferential  solvation. 
If they  are  hydrophilic,  
the  water-like   component 
is enriched near a  rod, even when the polymer  backbone 
is hydrophobic.   
In such complex situations,  the original concept of the 
Manning-Oosawa  condensation is not  
available (or at least  needs to be modified) 
to understand the counterion distribution. 
As a byproduct, we will predict   a first order 
phase transition between weakly 
 and strongly ionized states of a   rod.
%while the composition of the more polar component $\phi_B$  
%is small  far from it. 
Here the  ionization is assumed to be very 
weak for $\phi_B = 0$, but   increase 
with increasing $\phi_B$.  
Our prediction is that the progress of ionization 
can occur as a discontinuous change. 
If a polymer chain is in an expanded 
 state in the weakly ionized phase, 
it should be more expanded in the strongly ionized phase. 
It is analogous to the 
prewetting phase transition of fluids 
on a planar boundary wall \cite{Cahn,Ebner,Evans}.

In Subsec.IIA, 
we will present a Ginzburg-Landau theory for one component solvents  
whose minimization gives  $\alpha$. 
In Subsec.IIB, we will analyze $\alpha$  and 
the counterion density $n_1$ 
 on the basis of some exact relations.
In Subsec.IIIA, we will set up 
a Ginzburg-Landau model 
for two component solvents, which 
includes  $\alpha$, 
 $n_1$, and the composition $\phi$ 
as fluctuating variables.
In Subsec.IIIB, we will 
numerically  examine these quantities in equilibrium 
for various parameters  without salt. In particular, 
we will  treat 
a hydrophobic rod 
and hydrophilic counterions, 
where the dissociation 
sensitively depends on the ambient composition. 
In Appendix B, we will 
examine the ion distributions 
in one component solvent with salt. 
In Appendix C, we will present a simple 
 thermodynamic  theory of the 
 solvation shell formation at small 
 composition $\phi_B$ 
  of  the more polar  component.

\section{Rod in one-component solvent}

We consider an ionizable  
polymer chain with a  long persistence length 
 in a one-component solvent. 
Its shape  is a rod or an  expanded coil.   
Necklace-like globules\cite{Barrat,Levin,Volk,Holm,Rubinstein,Baigl1} 
 are outside the scope of this 
work,  which appear in poor solvents with increasing 
ionization.   Assuming low ion densities far from the rod, we 
 neglect the formation of dipole pairs 
and ion clustering  \cite{Levin} 
and the free energy contribution  from the charge density 
fluctuations ($\propto 
\kappa^{3}$ with $\kappa$ being  
the Debye-H$\ddot{\rm u}$ckel  
 wave number)  \cite{Levin,Muth,Onukibook}.

As a simple model \cite{Fuoss},  
a polymer chain  is treated 
as  a cylinder with radius $b$ and length $L\gg b$ 
The system is in a cylindrical 
cell with radius $R \gg b$. 
We   assume that  all the quantities  
depend only on the distance $r$ from the  center  
of the rod neglecting  the end effects.  
Here we   neglect   
the discreteness of the charges along the 
chain \cite{Me,Burak}.

\subsection{Ginzburg-Landau free energy}

Let the  density of ionizable  groups on a rod   be 
 $\sigma _0$  and the 
 degree  of ionization  be
$\alpha$ in the range $0\le \alpha\le 1$. 
We assume homogeneity along the rod. 
Each ionized group has charge $-e$ and the counterions 
are monovalent. The  charge density along the rod is   
$-e\sigma$ with 
\be 
\sigma=\sigma_0\alpha
\en 
per unit length. 
The mobile ions are distributed 
 in the region $b<r<R$ and $0<z<L$. Their 
 densities are written as 
$n_i(r)$ and their charges are   
$Z_ie$  ($i=1,2, \cdots)$. In 
this work  $n_1(r)$ denotes the density of  the counterions 
from the rod plus 
the cations of the same species added as a strong salt 
(see Appendix B). 
Then  $Z_1=1$.  The charge density in 
the region $b<r<R$ is written   as 
\be 
\rho= e\sum_{i}  Z_in_i,
\en 
where the summation  
is over all the mobile ions $i=1,2,\cdots$. 
We  assume the overall charge neutrality,  
\be 
\int d{\bi r}\rho (r) =Le\sigma,  
\en 
where 
$\int d{\bi r}(\cdots)= 2\pi L \int_b^R dr r(\cdots)$ is the 
integral in the cell outside  the rod. 
In the present one-component case 
 the dielectric constant 
$\ve$ is assumed to be a  constant. 
The  electric potential  $\Phi(r)$ satisfies 
\be 
\ve  \bigg (\frac{d^2}{dr^2}+ 
\frac{1}{r}\frac{d}{dr}\bigg)  \Phi= -4\pi \rho . 
\en 
We impose the boundary condition  
 on  the electric field $E(r) = -d\Phi(r)/dr$ as    
\be 
 E(R)=0 .   
\en 
That is,  there is no surface charge on the outer surface. 
See the review by Dobryinin and Rubinstein \cite{Rubinstein} 
for a theory in the case of nonvanishing $E(R)$. 
With  the aid of  Eqs.(2.3) and (2.5), integration 
of Eq.(2.4) in the region $b<r<R$ 
gives the boundary condition, 
\be 
E(b)=-2e\sigma/\ve  b.
\en  
Our theory remains invariant 
with  respect to the shift 
of the potential $\Phi(r) \to \Phi(r)+$const.

Hereafter we set the Boltzmann constant equal to  unity. 
The Helmholtz  free energy  $F$ of our system  is divided in 
 two parts as $F=F_0+F_d$, where  $F_0$  
consists of  the entropic and 
electrostatic contributions and $F_d$  is the free energy of 
dissociation. They are written as 
\cite{Bu1,Joanny,Bu2,Muth,Burak,Onuki-Okamoto}  
\begin{eqnarray}
&& \hspace{-1cm} \frac{F_0}{T} = \int d{\bi r} 
\bigg [\sum_{i} n_i (\ln (n_iv_{0} )-1)   
+\frac{\ve {E}^2}{8\pi T}  \bigg], \\
&&\hspace{-1cm} 
\frac{F_d}{T} = L\sigma_0 [ 
\alpha\ln\alpha+ (1-\alpha)\ln(1-\alpha)]
+ L\Delta_0 \sigma . 
\end{eqnarray} 
In $F_0$,  the volume   $v_{0}$ 
 is taken to be independent of $i$, 
which is allowable without loss of generality  \cite{v0}. 
In $F_d$,  $L\sigma_0$ is  the total number of the ionizable 
groups and $\Delta_0$ is  
the  energy needed for ionization (see Eq.(2.14) for  
the dissociation constant in terms of  $\Delta_0$).  
The free ions   can interact with polar segments on a chain, 
but such an interaction is neglected.

In equilibrium we minimize 
the grand potential $\Omega= 
F-\int d{\bi r} \sum_i\mu_i n_i$ 
under the charge neutrality condition Eq.(2.3), 
where $\mu_i$ ($i=1,2,\cdots$) are 
appropriate constant chemical potentials.  
We  thus minimize  
\be 
\Omega= F-\int d{\bi r} \sum_i 
(\mu_i- T\lambda Z_i)  n_i-  T\lambda L \sigma
\en  
with respect to $\alpha$ and  
$n_i(r)$, where 
 $T\lambda$ is   the  Lagrange 
multiplier.   However, if 
we may replace  $\Delta_0$  by 
$\Delta_0-\mu_1/T$  in $F$, we 
 obtain  $\Omega=F$ without salt.

The electrostatic 
energy $ F_e\equiv  \int d{\bi r}
{\ve {E}^2}/{8\pi }$ changes  
with respect to small 
variations  of $\sigma$, $n_i$, 
and $\ve$  as \cite{OnukiPRE}     
\be 
\delta F_e 
=\int d{\bi r} \bigg [
e \Phi \sum_{i}Z_i \delta n_i - 
\frac{E^2}{8\pi}\delta\ve\bigg]  -e\Phi(b)\delta\sigma , 
\en 
where  use has been made of 
Eqs.(2.4)-(2.6).  Here 
$\delta\ve=0$ for   one component 
solvents, but $\delta\ve$ will be 
nonvanishing for two component solvents.
The minimum conditions $\p \Omega/\p \alpha=\delta 
\Omega/\delta n_i=0$  are rewritten as   
\bea
\frac{\alpha}{1-\alpha} &=& 
\exp\left[\frac{e\Phi(b)}{T}+\lambda-\Delta_0\right],
\\
n_i(r)&=&     n_{0i}\exp\left[-Z_i \left(\frac{e\Phi(r)}{T}+\lambda\right) \right]  , 
\ena
where  $n_{0i}=v_0^{-1}\exp({\mu_i/T})$  
are constants.  We notice that $e\Phi/T$ and $\lambda$ appear  in the sum 
$e\Phi/T+\lambda$, so we are allowed 
to set $\lambda=0$  by redefining 
 $e\Phi /T+\lambda$ as $e\Phi /T$. Substitution of Eq.(2.12) into Eq.(2.4)  
yields the 
Poisson-Boltzmann equation.

From Eq.(2.12)  we have the relation $\sum_i 
[T\ln (n_iv_0)- \mu_i+ T\lambda Z_i] n_i= -\rho \Phi$. 
Further noting the relation 
$\int d{\bi r}\rho(r)[\Phi(r)-\Phi(b)]
=2F_e$ and 
using Eq.(2.11) we find that 
 $\Omega$ assumes a negative 
minimum  expressed as 
%\be 
%\frac{\Omega}{TL }= -2\pi\int_b^R 
% dr r\bigg[\sum_i n_i + 
%\frac{\ve E^2}{8\pi T}\bigg]+\sigma_0\ln(1-\alpha),
%\en                             
\be 
\frac{\Omega}{TL }= -\sigma +\sigma_0\ln(1-\alpha) 
 -\frac{F_e}{TL}- \frac{N_{\rm s}}{L},
\en                             
where  
$N_{\rm s}= \int d{\bi r}\sum_i n_i -\sigma L$ 
is the  number of the ions added as a salt.  
In Eqs.(2.28) and (2.29) below, 
  $\Omega$ will be calculated  explicitly 
 without salt ($N_{\rm s}=0$).

As $r \to b$, we have $n_1(b)= n_{01}\exp[-e\Phi(b)/T-\lambda] $ 
for monovalent  counterions. 
We thus obtain the mass action law 
(equation of ionization    equilibrium), 
\be 
\frac{\alpha}{1-\alpha}n_1(b)
 =K_0 ,
\en   
where $K_0$ is the dissociation constant defined by  
\be 
K_0  =  n_{01}e^{-\Delta_0}= 
v_{0}^{-1}e^{\mu_1/T -\Delta_0}. 
\en  
Since  Eq.(2.14) yields 
 $\alpha=1/[1+n_1(b)/K_0]$, $\alpha$ increases   
up to unity for $n_1(b)\ll K_0$. 
Note that $K_0 $ is a measurable quantity 
and should be invariant with respect to the choice of $v_0$ 
in Eq.(2.7) \cite{v0}.  
The mass action law of chemical reaction  
is often expressed  in terms of the 
pH of the solution 
if the cations are  protons \cite{Bu2}. 
In our case Eq.(2.14)   holds 
for the counterion density 
 $n_1(b)$ at the polymer backbone.

\subsection{Analysis in the salt-free case}

The Poisson-Boltzmann 
equation  can be solved exactly  without salt 
 under the boundary conditions Eqs.(2.5) and (2.6)\cite{Fuoss, An}. 
The solution  is parameterized by 
the normalized line density $\ell_B \sigma$, 
called the Manning parameter, 
and the radius ratio $R/b$.  Here   
$
\ell_B=e^2/\ve T
$ 
is the Bjerrum length. 
We examine how the degree of ionization   
$\alpha$ is determined for 
each given $\ell_B \sigma_0$.  
In this subsection we will  give 
approximate results for large $M$. 
Some  exact results are summarized in Appendix A.
Furthermore, we will discuss the 
effect of salt in Appendix B.

\subsubsection{Counterion density  
and degree of ionization}

It is known that 
$n_1(r)\propto r^{-2}$ and $E(r)\propto r^{-1}$ 
roughly hold for $r\gg b$. 
It is convenient to introduce  the dimensionless 
function,
\be  
{F_1(r,\sigma)} = 2\pi \ell_B r^2 n_1(r) ,
\en 
which  behaves differently depending on whether 
$\sigma< \sigma^*$ or $\sigma> \sigma^*$ with 
$\sigma^*$ being  a critical line density, 
\be 
\sigma^*= \ell_B^{-1}/(1+M^{-1}).  
\en 
Here we define the dimensionless   parameter, 
\be 
M=\ln (R/b),
\en 
which is assumed to be considerably larger than unity.

%The normalized  electric  potential (2.11) 
%is expressed as 
% $U(r)=2\ln(r/b)-\ln F_1(r,\sigma)+$
%const. 

% 1
\begin{figure}
%[htbp]
\begin{center}
\includegraphics[scale=0.5]{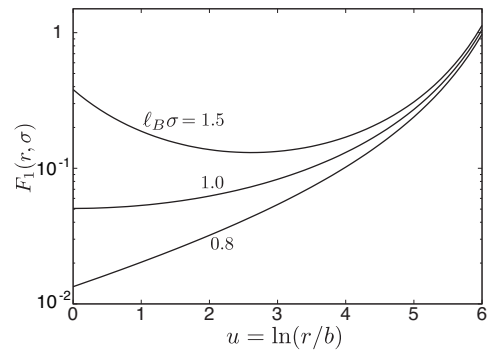}
\caption{
$F_1(r,\sigma )= 2\pi \ell_Br^2n_1(r)$ 
vs $\ln(r/b)$ on  a logarithmic 
scale for $\ell_B\sigma=0.8,1$, and 1.5  
with $M=6$ for 
one component solvent without salt. 
%The corresponding values of $B$ are 
%$0.163$, $0.225$, and $0.361$, respectively. 
The slope of the curve at $r/b=1$ 
changes its sign at $\ell_B\sigma=1$. 
}
\end{center}
\end{figure}

In Fig,1, we show $F_1(r,\sigma)$ 
for $\ell_B\sigma=0.8,1$, and 1.5 at $M=6$ 
treating  $F_1$  as a  function of 
 $u\equiv \ln(r/b)$. The area below each curve  is the 
 Manning parameter since   
\be 
 \int_0^M du 
F_1(r,\sigma)=\ell_B \sigma.
\en 
We notice that $F_1(r,\sigma)$   changes 
 most drastically  at $r=b$ and  most weakly at $r=R$.   
Its value at $r=b$ is 
  the normalized  counterion density at the rod surface 
since    $F_1(b,\sigma)= 2\pi \ell_Bb^2n_1(b)$. 
To examine it, we introduce  a scaling variable  $q$  by 
\be 
q= (\ell_B\sigma-1)M.  
\en 
At $\sigma=\sigma^*$ we have 
$q=-M/(1+M)\cong -1$.  The relations in Appendix A 
yield 
\bea
F_1(b,\sigma) 
&\cong& 2 \sigma \ell _B(\sigma\ell_B-1)^2 e^{2q} 
\quad (-q\gg 1) ,\nonumber\\
&\cong & (\sigma\ell_B-1)^2 \quad (q \gg 1) . 
\ena 
For $|q| \sim 1$ we have   $F_1(b,\sigma)
\sim M^{-2} $. Thus $F_1(b,\sigma) $ is  
very small for  $-q\gs 1$ with increasing $M$, 
while it tends to 
be independent of $M$ for $q\gg 1$.

% 3
\begin{figure}
%[htbp]
\begin{center}
\includegraphics[scale=0.42]{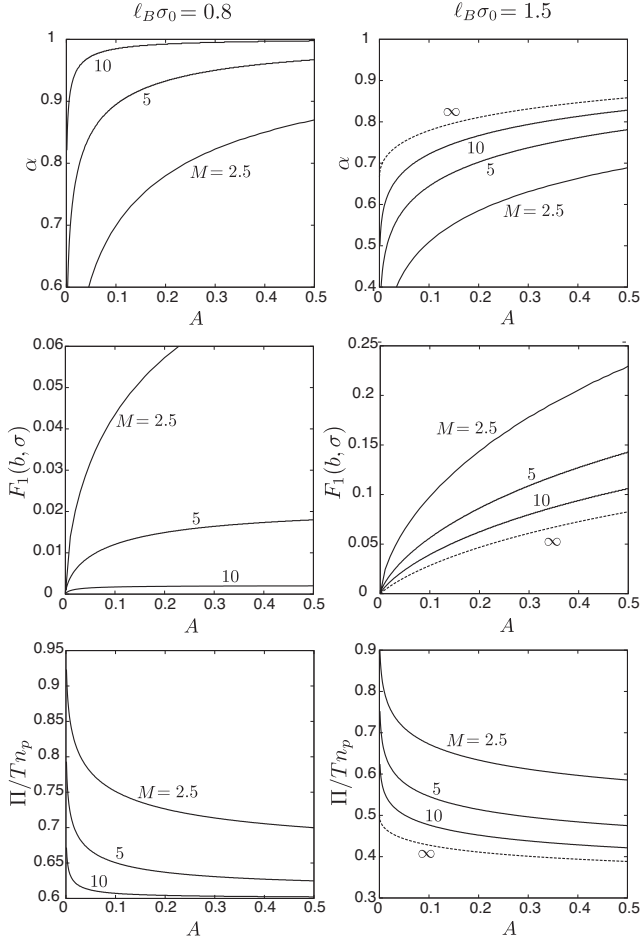}
\caption{Degree of ionization $\alpha$ (top) 
,  normalized counterion density 
$F_1(b, \sigma )$ on the rod surface (middle), and 
 osmotic pressure $\Pi$ at $r=R$ 
divided by 
$Tn_p= T\sigma/\pi R^2$ (bottom) 
 as functions of $A$ in Eq.(2.22) for $\ell_B\sigma_0=0.8$ (left) 
and 1.5 (right) in one component solvent  without salt.  }
\end{center}
\end{figure}

%4
\begin{figure}[htbp]
\begin{center}
\includegraphics[scale=0.5]{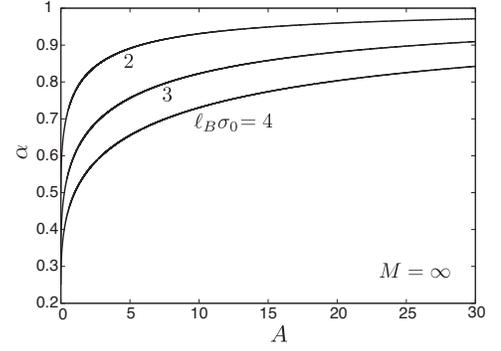}
\caption{Degree of ionization 
 $\alpha$  vs  $A$   for 
$\ell_B\sigma_0=2$, 3, and 4 obtained from the asymptotic 
equation (2.26) 
 in the limit  $R \to \infty$ in one component solvent without salt. 
Here $\alpha \to 1/\ell_B\sigma_0$ as  $A\to 0$, 
while $\alpha \to 1$ for  $A\gg 1$.}
\end{center}
\end{figure}

We now calculate the degree of ionization 
 $\alpha$ from the mass action 
law Eq.(2.14).  We introduce 
a  normalized dissociation constant, 
\be 
A=2\pi \ell_B b^2 K_0.  
\en  
In terms of $A$,   Eq.(2.14) is rewritten as 
\be 
\alpha F_1(b,\sigma)= (1-\alpha)A.
\en   
In  Fig. 2, we plot   numerical results of $\alpha$,  
$F_1(b, \sigma )$,  and $\Pi /Tn_p$ 
for $\ell_B\sigma_0=0.8$ (left) 
and  $1.5$ (right). 
Here we introduce  the osmotic pressure at $r=R$,  
\be 
\Pi=T n_1(R)
\en 
where the Maxwell stress vanishes from 
Eq.(2.5).  The   $n_p= \sigma/\pi R^2$ 
is the counterion density 
for the  uniform  distribution.

The first line of Eq.(2.21) gives 
\be 
(1-\ell_B \sigma_0 \alpha)^2 \alpha ^2/ (1-\alpha ) \cong  
{A} 
(R/b)^{2q}/{2\ell _B \sigma _0},      
\en 
for $-q= (1-\ell_B\sigma )M \gg 1$. 
The right hand side anomalously depends on $R$ 
and is very small 
for $R/b\gg 1$. 
On the other hand, the second line  of Eq.(2.21) gives 
\be
(\ell_B \sigma_0 \alpha-1)^2 \alpha \cong  A (1-\alpha ).    
\en
for $q= (\ell_B\sigma-1)M \gg 1$. This cubic 
 equation of $\alpha$ 
is  the asymptotic equation  independent of $R$, 
yielding  a unique solution in the range 
$( \ell_B \sigma_0)^{-1}<\alpha<1$. 
Namely,  
$\alpha \rightarrow (\ell_B\sigma_0)^{-1}<1$ 
as $A\rightarrow 0$, while  
$\alpha \rightarrow 1$ with increasing $A$. 
In Fig. 3,  we show this  limiting 
$\alpha$ obtained from  Eq.(2.26) 
as a function of $A$ for $\ell_B\sigma_0>1$, 
where the increase of $\alpha$ from 0 to 
$(\ell_B\sigma_0)^{-1} $ in the 
narrow region $0<A\ls (b/R)^2$ 
is not shown.

\subsubsection{Grand potential   and  effective polymer-solvent 
interaction}

In equilibrium without salt, 
the grand potential  
$\Omega$ in  Eq.(2.13) 
consists of three negative parts as 
\be 
{\Omega}/{TL} =  -2\sigma
 -\ell_B^{-1} {\cal F}_e(\ell_B\sigma) 
+ \sigma_0 \ln (1-\alpha ) , 
\en  
The electrostatic energy 
$F_e=\int d{\bi r}\ve E^2/8\pi$ 
is equal to $TL(\sigma + \ell_B^{-1} {\cal F}_e)$.  
See Appendix A for the exact expression for the scaling 
function ${\cal F}_e(s)$, which was derived by 
Naji and Netz \cite{Netz}.   
(i) For small $\ell_B\sigma\ll M^{-1}$ 
the electrostatic part is negligible 
and ${\Omega}/TL  \cong -\sigma+ \sigma_0 \ln (1-\alpha )  $. 
(ii) For  $M^{-1}\ll  \ell_B\sigma\ll 1$ 
we have  
\bea 
&&{F}_e\cong TL \ell_B \sigma^2M, \nonumber\\ 
&& {\Omega}/TL  \cong -\ell_B \sigma^2
M + \sigma_0 \ln (1-\alpha ).
\ena  
(iii) On the other hand, 
for $\ell_B\sigma-1\gg M^{-1}$ and $\ell_B \sigma \ll M$, 
we find 
\bea 
&&{F}_e \cong TL\ell_B^{-1}M,\nonumber\\
&&{\Omega}/TL  \cong  -\ell_B^{-1}M+\sigma_0 \ln (1-\alpha ). 
\ena 
Here the electrostatic energy is nearly equal to that 
of a  charged rod  
at $\sigma=\ell_B^{-1}$ without screening. 
This result suggests 
 that  a fraction of $1-(\ell_B\sigma)^{-1}$ 
of the  counterions are 
localized around  the rod, 
as well as the osmotic pressure behavior 
in Eq.(A4).

Polymer chains in water  are often hydrophobic 
without ionization but can be 
hydrophilic  with ionization \cite{Volk,Rubinstein,Baigl1}. 
This means that  the  polymer-solvent 
interaction parameter, written as $\chi_{ps}$, 
is effectively decreased upon ionization. 
Its decrease  $\Delta\chi_{ps}$ may be defined  as follows. 
In the Flory-Huggins free energy density of polymer solutions, 
the interaction part  
is written as $Tv_0^{-1}\chi_{\rm ps}\phi_p$ 
for small polymer volume 
fraction $\phi_p$, where 
 $v_0$ is  the solvent volume.  We map our polyelectrolytes 
system to a neutral polymer system. 
In the present  case,  the polymer volume is 
$\pi b^2L$ in the total volume $V=\pi LR^2$ so that 
$\phi_p=  (b/R)^2$. We set 
\be 
T v_0^{-1}  \Delta\chi_{ps}\phi_p= \Omega/V + \Pi .
\en 
Without salt we have $\Pi=Tn_1(R)$ and 
\be 
\Delta\chi_{ps}=
 \frac{v_0}{\pi b^2}\bigg[\frac{\Omega}{TL}+ 
\pi R^2 n_1(R)
\bigg]. 
\en  
Here in the brackets, the first term 
is  given by Eq.(2.27) 
and the second term is equal to 
$F_1(R,\sigma)/{2\ell_B}$ from Eq.(2.16). 
We confirm the negativity of $\Delta\chi_{ps}$ 
as follows. 
(i) For small $\ell_B\sigma\ll M^{-1}$ 
we have $\Delta\chi_{ps} \cong 
(v_0 /\pi b^2 ) \sigma_0\ln (1-\alpha)$. 
(ii) For  $M^{-1}\ll  \ell_B\sigma\ll 1$ 
we have  $\Delta\chi_{ps} \cong 
(v_0 /\pi b^2 ) [-\ell_B\sigma^2+
\sigma_0\ln (1-\alpha)]$. 
(iii) For 
$\ell_B\sigma-1\gg M^{-1}$ and $\ell_B \sigma \ll M$, 
we have  $\Delta\chi_{ps} \cong  (v_0 /\pi b^2)
\Omega/TL$, where  $\Omega$ is given by Eq.(2.29).

\section{Rod in two component solvent}
\setcounter{equation}{0}

We next consider a charged rod 
in a nearly incompressible two component  
solvent such as a  mixture of water+alcohol or  water+ 
 organic solvent in the  same geometry 
 as in the one component case. 
The volume fraction of the water-like component  
is  $\phi(r)$ and that of the less polar  component is 
$1-\phi(r)$.  
The molecular volumes of the two components 
(the inverse densities of the pure components) 
are assumed to be given by a common volume $v_0=a^3$, which 
 may be  equated with   $v_0$ in Eq.(2.7). Hence  
  $\phi$ will  also be called the composition. 
We assume  small  ions and 
neglect their  volume fraction.   
As in the previous section,  
 $n_1(r)$ is the density of   the counterions 
plus  the cations of the same species added as a salt.

We suppose a polymer chain with a hydrophobic backbone 
(like polystyrene) and ionizable groups attached to it 
\cite{Rubinstein}.  
%Its ionization increases with addition 
%of  the water-like component.  
% $\alpha\ll 1$ for $\phi=0$, 
% $\alpha$ increases with increasing $\phi$.
With addition of water, 
we first need to consider  the 
formation of solvation  (or hydration)  shells 
 composed of several water molecules 
around hydrophilic charged particles \cite{Is}.   
This can occur even at very small 
water content  \cite{Osakai}, 
as will be discussed  in  Appendix C.

\subsection{Ginzburg-Landau theory}
\setcounter{equation}{0}

For not very small  bulk water composition $\phi_B$, 
the water composition $\phi(r)$ 
varies  on mesoscopic scales. 
To describe such situations, we present  a Ginzburg-Landau 
theory with the gradient free energy 
of the composition \cite{Onukibook}.

\subsubsection{Free energy including solvation interaction}

The total free energy  is composed 
of three parts as $F=F_0+F_d+\Delta F$, where 
 $F_0$ is given in Eq.(2.7) and $F_d$ in Eq.(2.8). 
Notice that  $\Delta_0$ in Eq.(2.7) 
should be replaced by $\tilde{\Delta}_0$ in Eq.(C7) 
to account  for the solvation  
shell formation. 
Hereafter  we will redefine   $\Delta_0$  
 as the right hand side of 
Eq.(C7). Then  $T\Delta_0$ in the 
following is  the renormalized 
dissociation energy per counterion.

Assuming the homogeneity along the rod, 
we write the additional  contribution $\Delta F$ in the form,       
\bea 
\frac{\Delta F}{T} &=& \int d{\bi r}\bigg[ 
\frac{f_0(\phi)}{T}  + \frac{C}{2}|\nabla \phi|^2- 
 \sum_{i} g_i {n_i} \phi\bigg] \nonumber\\
&& -2\pi b L  \gamma  \phi_s   -L  \Delta_1\sigma_0\alpha \phi_s .
\ena 
The space integral is in the cell, $b<r<R$ and $0<z<L$.   
The free energy density  $f_0(\phi) $ is taken to be     
the Bragg-Williams form,  
\be 
f_0 = \frac{T}{v_0} [
 {\phi}  \ln\phi + (1-\phi)\ln (1-\phi) 
+ \chi \phi (1-\phi) ], 
\en 
where $\chi$ is the interaction parameter 
 dependent  on $T$ 
and its  mean-field critical value is $2$   
in the absence of ions. 
The  coefficient $C$ of the gradient 
part is a positive constant of order $a^{-1}$ 
and will be taken to be $3a^{-1}$ in our numerical analysis. 
The   coupling terms $(\propto g_i$)  
arise from the  ion-dipole  interactions among 
the ions  and  the polar solvent molecules.  
In the second line of Eq.(3.1), 
the   term proportional to $\gamma$  arises 
 from the short-range interaction between the rod surface 
and the solvent molecules \cite{Cahn}, 
while the term proportional to $\Delta_1$  represents 
the solvation interaction between the surface charge 
and the solvent  molecules. We  neglect  
the interaction between the mobile ions 
and the uncharged monomers (which can be important for polar  monomers 
 \cite{Onuki-Okamoto}). 
Hereafter,   
\be 
\phi_s=\phi(b)
\en 
is the surface value of the composition 
(outside the solvation shells under the condition (C8)).   
 These molecular interactions of the rod 
are characterized by 
the two constants   $\gamma$ and $\Delta_1$, 
which can be either negative or positive.

 In $F_0$ in Eq.(2.7), 
 the dielectric constant $\ve$ 
is  assumed to depend  on the composition 
in   the linear form \cite{Debye-Kleboth},   
\be 
\ve(\phi)=\ve_0  + \ve_1 \phi ,
\en 
where $\ve_0+\ve_1$ 
is the dielectric constant of the first component and 
$\ve_0$  is that of the second component. 
Thus $\ve(\phi)$ is inhomogeneous.
 The electric potential
 $\Phi(r)$ satisfies
\be
\frac{1}{r} \frac{d}{dr} r\ve (\phi)\frac{d}{dr} \Phi=-4\pi \rho, 
\en
under  the boundary conditions $E(R)=0$ and   
\be 
E(b)=-2e\sigma/\ve (\phi _s) b.
\en 
%and  for the electric field $E(r) =-d\Phi/dr$.
%Using $\sigma$ we  integrate the Poisson equation (3.5) as 
%\be
%\frac{d\Phi(r)}{dr}
% =\frac{2}{{r} \ve (\phi) } 
%\left[e \sigma  -2\pi 
%\int _b^{{r}} dr'  r' \rho (r') \right] .
%\en

\subsubsection{Equilibrium  
and composition-dependent mass action law}

We minimize the grand potential 
$\Omega$ defined as in Eq.(2.9) with  
 $F=F_0+F_d+\Delta F$.  
The counterparts of Eqs.(2.11) and (2.12) read   
\bea
&&\frac{\alpha }{1-\alpha}= \exp \left[\lambda+ \frac{e\Phi(b)}{T} -\Delta _0+\Delta _1\phi _s \right], 
 \\
&&n_i(r) = n_{0i} \exp \left[-Z_i \frac{e\Phi(r)}{T} +g_i\phi(r) \right] ,
\ena 
where $n_{0i}=  v_{0} ^{-1}e^{\mu_i/T-Z_i\lambda}$. 
As in Sec.1, we may set $\lambda=0$ without loss of generality. 
From these equations the mass action law follows as  
\be 
\frac{\alpha}{1-\alpha}n_1(b)
 =K(\phi_s)  .
\en 
The  dissociation  constant 
$K(\phi_s)$  depends on  the surface composition 
$\phi_s$ as 
\be 
K(\phi_s)=  n_{01} e^{-\Delta_0+ 
(\Delta_1+ g_1)\phi_s}.
\en 
%See Eq.(C7) for  $\tilde{K}_0$. 
With increasing $\phi_s$, $K(\phi_s)$  
 increases (decreases) for positive (negative) 
$\Delta_1+ g_1$. We are not aware of experimental data 
on the composition dependence  of $\alpha$ for polymers  
in mixture solvents. On the other hand, 
strong composition dependence 
has been reported in 
dissociation of weak acids in 
various aqueous 
mixtures \cite{Ohtaki,Bonn}.

The functional derivative 
$h=\delta F/\delta \phi$ at fixed $n_i$ and $\alpha$ 
is  the chemical potential difference 
of the two components divided by $v_0$ 
and is homogeneous in equilibrium. 
From  Eqs.(2.10) and (3.1) we obtain 
\be 
h=  f'_{0} (\phi )-
TC\nabla ^2 \phi -T\sum _{i}g_in_i -
\frac{\ve _1}{8\pi }E^2 ,
\en 
where $f'_{0} (\phi )= \p f_{0} (\phi )/\p\phi$ and $E= -d\Phi/dr$. 
In equilibrium $h$ is a homogeneous constant.

With these results 
it is convenient to redefine the grand potential for  two component solvent as 
\be 
\Omega= F-\int d{\bi r} \bigg[\sum_i 
\mu_i  n_i +f_0(\phi_B)+ h(\phi-\phi_B) \bigg],
\en  
where $\phi_B=\phi(R)$ is the value of $\phi$ at $r=R$.  
As in Eq.(2.13) some calculations give                                  
\bea 
&&\frac{\Omega}{TL}
= 2\pi\int_b^R d{r}r \bigg[\frac{1}{T}\hat{f}_0 (\phi)  
+ \frac{C}{2}\phi'^2\bigg]-2\pi b \gamma \phi_s 
\nonumber\\
&&\hspace{5mm} -\sigma +\sigma_0\ln(1-\alpha)-\frac{F_e}{TL}
   -\frac{N_{\rm s}}{L}.
\ena
In the first line we write                                   
 $\phi'=d\phi/dr$ and define 
\be 
\hat{f}_0(\phi) = f_0(\phi)-f_0(\phi_B)- h(\phi-\phi_B). 
\en 
The first line of Eq.(3.13) gives  
the compositional  contribution. 
The second line is of the same form as the right hand side 
of Eq.(2.14) with $N_s$ being the number of ions added as  salt.

Small variations of $\phi$, $\alpha$, and $n_i$ 
 yield an  incremental change of  $\Omega$, 
written as  $\delta \Omega$.  
 With  the equilibrium  conditions (3.5), (3.7), (3.8), 
and (3.11),  its linear terms proportional 
to $\delta\alpha$ and $\delta n_i$ 
vanish. There remain  surface parts of the form,     
\bea 
\frac{\delta \Omega}{LT} 
&=& 2\pi C [(r\phi'\delta \phi)_{r=R}-
(r\phi'\delta \phi)_{r=b}]\nonumber\\ 
&&-(2\pi b\gamma+\Delta_1\sigma)\delta\phi(b).  
\ena 
We set $\delta \Omega=0$  
for any boundary variations 
$\delta\phi(b)$ and $\delta\phi(R)$ to   obtain 
\bea 
&&\phi'(R)=0,\\
&&C \phi'(b)= -\gamma-(\Delta_1/2\pi b)\sigma  ,
\ena 
which are the boundary conditions on  $\phi$ 
in solving the equation $h=$const. in Eq.(3.11). 
In our theory  the condition at $r=b$ in Eq.(3.17) depends on $\sigma$. 
%If $\gamma<0$, the first component is 
%repelled from the rod without ionization 
%(or the rod is hydrophobic). 
For positive (negative) 
$\gamma+\Delta_1 \sigma/2\pi b$, 
the first component tends to be  attracted to (repelled from) 
the rod.   A similar  boundary condition has been used 
 in the gradient theory of  the wetting transition 
 on a planar wall \cite{Cahn}.

We use the gradient free energy. 
It is usually  derived from the gradient expansion of the 
two body van der Waals interactions and is  justified near 
the critical point.  
In the following, however, 
 we will present  numerical results  in the 
composition range  $0.15<\phi_B<0.53$, 
imposing the boundary condition 
(3.17) in the gradient theory, 
where  $\phi_B$ is the composition far from the rod.  
When the rod radius $b$ is of order $a$ 
and $\phi$ changes steeply around the rod, 
a density functional theory without the gradient 
expansion should yield more reliable results \cite{Ebner,Evans}.

\subsubsection{Solvation interaction}

In the  original Born theory 
\cite{Born},  the solvation chemical 
potential was  given by 
$\mu_{\rm sol}^i(\phi) = Z_i^2 e^2/2\ve(\phi) R_i$ 
for ion species $i$,  
where $Z_ie$ is the charge, 
 $\ve(\phi)$ is the solvent dielectric constant 
dependent on $\phi$, 
and $R_i$ is a microscopic length 
called the  Born radius. 
However, this formula is a crude approximation 
and is not applicable to hydrophobic ions. 
In this paper we assume 
the form  
\be 
\mu_{\rm sol}^i(\phi) 
=\mu_{\rm sol}^i(0) - Tg_i\phi.
\en 
The solvation contribution to the free energy density 
is given by $\sum_i \mu_{\rm sol}^i(\phi)n_i$, 
yielding the terms proportional to $g_i$ 
on the right hand side of Eq.(3.1).   
For a mixture of  water and oil, for example,   
 $Tg_i$ is  the difference of 
the solvation chemical potential 
in oil   and that in water. Therefore,  
  $g_i>0$ for hydrophilic ions and  $g_i<0$ for hydrophobic  ions 
in aqueous mixture solvents.

The difference of $\mu_{\rm sol}^i$ 
between the  coexisting phases coincides with 
 the  Gibbs transfer energy  $\Delta G_{\rm tr}^i$  
(per ion) in electrochemistry 
\cite{Osakai,Hung}. It becomes  $Tg_i$ 
in strong segregation in our theory.  
Data of $\Delta G_{\rm tr}^i$  for   water+nitrobenzene 
 at $T \sim 300$K suggest         
 $g_i\sim 15$ for monovalent ions such as Na$^+$  
and   $g_i \sim -15$ for 
 tetrarphenylborate BPh$_4^{-}$. 
For water+alcohol, 
 $|g_i|$ should not be smaller, since 
the dielectric constant of nitrobenzene 
is 35 and that of alcohol is smaller (24 for ethanol). 
The solvation coupling is thus very strong in aqueous mixtures.

\subsubsection{Analysis of behavior of  composition deviation}

The composition $\phi(r)$ tends to a constant $\phi_B$ 
far from  the rod. We treat  $\phi_B$  as a control 
parameter.    With a salt added, this behavior is obvious 
 if the Debye length $k_D^{-1}$ is shorter than the system radius $R$ 
and longer than the correlation length $\xi$ of the composition.  
In such a case Eq.(3.11) gives 
$\phi(r)-\phi_B \propto e^{-k_D r}$ far from the rod.

Without  salt, however, the decay is   algebraic. 
To show it, we linearize  Eq.(3.11) with respect to the 
 composition deviation $\delta \phi= 
\phi-\phi_B$   to  obtain  
\be
(1-\xi ^2 \nabla^2)
 \delta \phi  =\chi_s
g_1n_1+ \frac{\chi_s\ve_1}{8\pi T}E^2,
\en 
where we assume $|\delta\phi| \ll \phi_B$ at any $r$. 
We define the correlation length   $\xi$  
and the susceptibility $\chi_s$   as  
\bea
\xi   &=&  [TC/f''_0(\phi_B)]^{1/2},\\
\chi_s &=& T/f''_0(\phi_B)= \xi^2/C,
\ena
where $f''_0(\phi)= T[1/\phi(1-\phi)-2\chi]/v_0$ from Eq.(3.11). 
In this paper,  we set 
$C =3/a$; then, $\chi_s  \sim a\xi^2$.
The length  $\xi$ remains of order $a$ 
away from the  criticality, while it grows near the criticality.

The two terms on the 
right hand side of Eq.(3.19) 
roughly behave as $r^{-2}$ 
from the results for one component solvent. 
That is, if $F_1=2\pi\ell_{B0} r^2n_1(r)$ 
in Eq.(2.16) (or in Eq.(3.29) below) 
is of order unity, the two terms  are  of order 
$\chi_s g_1/2\pi\ell_{Bb}r^2$ and  
$\chi_s \ve_1/2\pi\ve_0\ell_{Bb}r^2$, respectively. 
We define two Bjerrum lengths as
\be 
\ell_{B0}= e^2/\ve_0 T , \quad 
\ell_{Bb} = e^2/\ve_B T,
\en 
at $\ve=\ve_0$  and $\ve=\ve(\phi_B)=\ve_B$, respectively. 
Thus $\ell_{Bb}=\ell_{B0}/(1+\ve_1\phi_B/\ve_0)$. 
Furthermore,  we are interested in  the strong 
solvation case  $g_1\gg 1$ with 
 $\ve_1/\ve_0 \sim 1$, where 
the first term dominates 
over the second  in the right hand side of 
 Eq.(3.19). Then $\delta\phi(r)$ decays for $r-b \gs \xi$ as  
\be 
\delta\phi (r) 
%\sim (\chi_s g_1/2\pi\ell_{Bb})r^{-2}
\sim  (g_1a \xi^2 /2\pi\ell_{Bb})r^{-2}.
\en 
Here we are assuming $|\delta\phi|\ls \phi_B$, 
which holds even for $r\sim \xi$ if 
$g_1a  /2\pi\ell_{Bb}\ls 1$.

\subsubsection{Effective charge 
density and modified Manning  law}

For large $r$ far from the rod, 
$\phi (r)$ tends to a constant $\phi_B$ 
and   $n_1(r)$ 
obeys the nonlinear Poisson-Boltzmann equation 
even in the salt-free case.   
We  introduce  a characteristic 
length $r_c$ in the range $b<r_c\ll R$. 
For  $r$ larger than  $r_c$, 
we may set $\ve(\phi)\cong \ve_B$ 
 with $\ve_B=\ve(\phi_B)$ and 
the potential $\Phi(r)$ 
obeys      
\be
-\ve_B\nabla ^2 \Phi \cong 
4\pi e n_0 \exp (-e\Phi/T) , 
\en 
where we set  $Z_1=1$  and 
$n_0=n_{01}e^{-\lambda}$. We  compare 
$\Phi(r)$ and $n_1(r)$  in our system 
approximately obeying Eq.(3.24) far from the rod 
and those  obtained as the solution 
of the nonlinear Poisson-Boltzmann equation,  
\be 
-\ve_B\nabla ^2 \Phi_{\rm PB}=  
4\pi e n_1^{\rm PB}, 
\en  
where $n_1^{\rm PB}(r)=n_0 \exp (-e\Phi_{\rm PB}(r) /T)$ is the 
corresponding counterion density.  
We set $\Phi_{\rm PB}'(R)=0$ at $r=R$ 
and  impose the boundary condition  at $r=r_c$ as  
\be 
\Phi_{\rm PB}'(r_c )=\Phi'(r_c), 
\en 
where $\Phi_{\rm PB}'(r)= d\Phi_{\rm PB}(r)/dr$ and 
$\Phi'(r)= d\Phi(r)/dr$.  From Eq.(3.25) 
 $ \Phi(r)\cong \Phi_{\rm PB}(r)$ 
and $n_{1}(r) \cong n_{1}^{\rm PB}(r)$ far from  the rod 
 $r>r_c$, but  significant  differences can arise 
in the region $r<r_c$.

In terms of $\Phi_{\rm PB}$ we define  
the effective charge density 
$\sigma_{\rm eff}$ on the rod by  
\be 
\sigma_{\rm{eff}}  =\ve_B b  \Phi_{\rm PB}'(b)/2e. 
\en 
From Eq.(2.6)  the real charge density $\sigma$ 
is given by $\sigma =\ve(\phi_s) b  \Phi'(b)/2e$.  
From Eqs.(3.5), (3.24),  and (3.25) 
the apparent  excess charge density 
$\Delta\sigma=\sigma-\sigma_{\rm{eff}}$ 
on the  rod is expressed as
\bea
\Delta \sigma&=& 2\pi n_0 
\int _b^{r_c}dr r [e^{-e\Phi(r)/T}-  e^{-e\Phi_{\rm PB}(r)/T}
]\nonumber\\
&=& \frac{1}{\ell_{B0}}\int_0^{u_c}du 
[F_1(u)- F_1^{\rm PB}(u)].
\ena 
In the  second line,  we set 
 $u=\ln(r/b)$ with  $u_c=\ln(r_c/b)$ being the upper bound 
and  define 
\bea 
&&F_1= 2\pi r^2\ell_{B0}n_1, \nonumber\\ 
&&F_1^{\rm PB}= 
2\pi r^2\ell_{B0}n_1^{\rm PB},
\ena 
as in Eq.(2.16).  These quantities are  functions 
of $u$. In the bottom plates of Fig.4, we shall see that $u_c$ 
may be pushed to infinity 
since the integrand in the second line of Eq.(3.28) 
vanishes   for large $u$. 
Here the effective Manning  parameter is 
$\sigma_{\mathrm{eff}}\ell_{Bb}$, 
where $\ell_{Bb}$ is defined in Eq.(3.22). 
In terms of 
$n_p^{\mathrm{eff}}\equiv \sigma _{\mathrm{eff}}/\pi R^2$
the asymptotic law for the osmotic pressure $\Pi$ 
in Eq.(A4) is changed  to 
\bea 
{\Pi} &\cong& Tn_p^{\mathrm{eff}} (1-\ell_{Bb}\sigma_{\mathrm{eff}}/2) \quad (\ell_{Bb}\sigma_{\mathrm{eff}} <1)
\nonumber\\
&\cong&  Tn_p^{\mathrm{eff}} /2\ell_{Bb}\sigma_{\mathrm{eff}}  \qquad (\ell_{Bb}\sigma_{\mathrm{eff}} >1). 
\ena

\subsubsection{Strong 
deformations of the counterion distribution}

We examine  the conditions  
of strong attraction of the counterions  due to 
a composition change 
around the rod. 
We assume  that $\phi_B$ 
is not very small and $\xi$ is not much separated from 
$ a$ for simplicity.  
If $g_1\gg 1$,  $ \delta \phi$ is  of 
order $g_1a /2\pi\ell_{Bb}$ at  $r \sim \xi$ from Eq.(3.23). 
Since $n_1 \propto e^{g_1\phi}$ from Eq.(3.8), 
an appreciable 
attraction of the counterions is induced due to 
the composition change when  
\be 
g_1^2a/2\pi \ell_{Bb} \gs 1. 
\en 
For large $g_1\gg 1$ the above 
 condition can be realized while 
$\delta\phi$ remains  small $(\ll \phi_B$).

Also  $\phi'(b)$ 
satisfies the boundary condition (3.17). 
This yields  a contribution to $\delta\phi(r)$, written as 
$(\delta\phi)_b(r)$. It   obeys 
$(1-\xi^2\nabla^2)(\delta\phi)_b=0$ 
in the bulk, so it  decays 
 rapidly far from the rod. 
If $\xi\ls b$, 
it is approximately written  as   
\be
(\delta \phi)_b (r)  
\cong \bigg
 [\gamma+ 
\frac{\Delta _1\sigma }{2\pi b}  \bigg]\frac{\xi}{C} 
\exp \bigg [-\frac{r-b}{\xi}\bigg ] .
\en
If the above composition change is positive, 
the counterions are significantly attracted to 
the rod  when 
\be 
g_1(\gamma+ {\Delta _1\sigma }/{2\pi b})   
{\xi}/{C}\gs 1.
\en 
If it is negative and its absolute value exceeds unity 
and if the condition (3.31) does not hold, 
the  counterions should  be 
repelled from the rod.

The criterions  (3.31) and (3.33) are  crude 
ones based on many assumptions. 
In particular,  $F_1(b)\sim 1$ has been assumed 
in deriving Eq.(3.31) and 
 the unknown $\sigma=\sigma_0\alpha$ 
is  contained in Eq.(3.33).
Setting up general criterions  is  at present 
difficult,  because many of the parameters strongly  affect the 
counterion distribution.  
Nevertheless, we recognize that the counterion distribution 
can be changed dramatically 
even for a slight change of the composition 
in the strong solvation condition $|g_1|\gg 1$.

\subsection{Numerical results 
for two component solvent   without  salt}

We present some numerical results for 
  a water-oil solvent without salt.  
In all the examples to follow, 
we numerically solve Eqs.(3.5), (3.7), (3.8),  and (3.11) 
under given boundary conditions   
by setting  
\bea 
&&b=2a,  
~ R/b=e^6\cong403, \nonumber\\ 
&&\sigma_0 =\pi /2a, ~ C=3/a,~\Delta_1=7,  
 \nonumber\\ 
&&{\ve_1}=2{\ve_0}, ~ \ell_{B0}=  
12a/\pi=6/\sigma_0,\nonumber
\ena 
where $\ell_{B0}$ is defined in Eq.(3.22). 
The solvation parameter $g_1$ will be 
either of  5, 7, or  10. Note that  $g_1$ can be 
larger in  real  situations, as discussed  below 
Eq.(3.18).  
The parameter $A=2\pi \ell_{B0}b^2K_0$  in Eq.(2.22)  
will be chosen to be small.  As a result,  little  ionization 
 occurs  without composition variations  
near the rod.  In our simulation the counterion density 
far from the rod is so small such that 
 the  solvent phase diagram is unchanged 
\cite{OnukiPRE}.

\subsubsection{Attraction of water  and 
counterions to a hydrophobic rod}

%7
\begin{figure}[htbp]
\begin{center}
\includegraphics[scale=0.42]{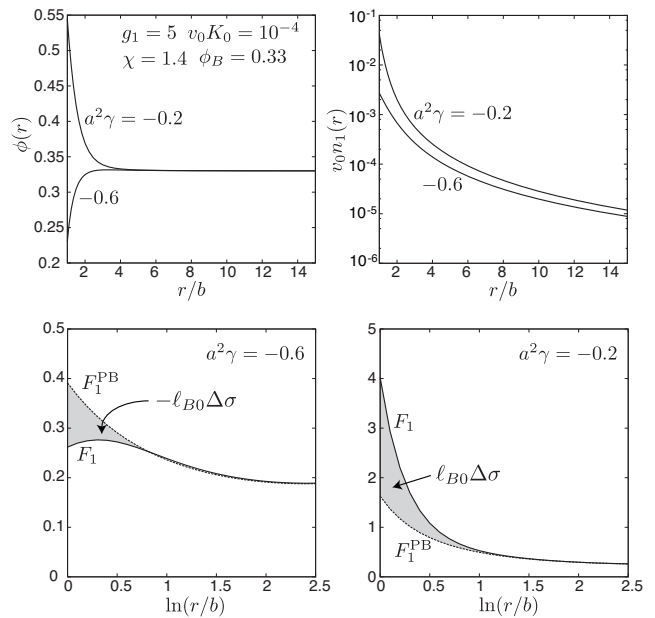}
\caption{Numerical results for a mixture solvent 
in the salt-free case for $\chi=1.4$, 
$\phi_B=1/3$,  $g_1=5$, and 
$A= 0.0096$ 
(on the point $(+)$ in  Fig.6). 
  Top plates: 
composition $\phi(r)$ (left) 
and  normalized counterion density $v_0n_1(r)$ (right)  
for $a^2\gamma=-0.6$ and $-0.2$. 
Bottom plates: $F_1(u)$ and $F_1^{\rm PB}(u)$ 
in Eq.(3.29)  vs $u=\ln(r/b)$ 
for $a^2\gamma=-0.6$ (right) 
and $-0.2$ (left).  The area between  these 
two functions (shaded) 
is  equal to the normalized excess charge 
density $\ell_{B0}\Delta\sigma$ 
from Eq.(3.28).}
\end{center}
\end{figure}

The interaction  between the solvent and 
 the rod in Eq.(3.1) leads  
to  the boundary condition (3.17) 
in our gradient theory. For $\gamma<0$ ($\gamma>0$),  
the  rod repels (attracts) 
 the water component and is hydrophobic (hydrophilic) 
without ionization. 
Even for $\gamma<0$, however, 
the right hand side of Eq.(3.17) changes from positive to 
negative with increasing the degree of ionization 
$\alpha$   and 
the condition (3.33) can eventually  hold  for 
$|\gamma|<\Delta_1\sigma/2\pi b$. 
Then the rod becomes  effectively hydrophilic. To examine the resultant 
preferential adsorption of the water-like 
component,  we  introduce 
\be
\Gamma= 
{2\pi} \int_b^R  dr r [\phi (r)-\phi _B], 
\en 
where $\phi_B$ is the value of $\phi(r)$ at $r=R$.

 Figure 4 illustrates 
 such a changeover in the one phase region 
for  (a) $a^2\gamma=-0.6$ 
and (b) $a^2\gamma= -0.2$.  We set 
 $\chi=1.4$,  $\phi_B=1/3$,  $g_1=5$,   
   and $v_0K_0=10^{-4}$. Then   
  $A=2\pi \ell_{B0}b^2K_0= 
0.0096$ from Eq.(2.22), $\xi=1.32a$ from Eq.(3.20), 
and $v_0K=10^{-4}e^{12\phi_s}$ from Eq.(3.10). 
In the upper left panel of Fig.4,  we present  
the  composition  profile $\phi(r)$ 
near the rod, which 
is repelled for (a) and attracted for  (b). 
However, as shown in Eq.(3.23), 
$\delta\phi=\phi-\phi_B$ has a positive 
tail ($r^{-2}$) without salt, giving rise to 
a positive logarithmic contribution 
$\sim (g_1a\xi^2/\ell_{B0})\ln (R/\xi)$ to 
the integral $\Gamma$ in Eq.(3.34). 
Due to this singular contribution 
we obtain   $\Gamma=0.706a^2$ 
for  (a), while  $\Gamma=3.99a^2$ 
for   (b). 
In  the upper right  panel, the  difference  
of $n_1(r)$ between the two cases 
 is large and  $n_1(r)$ is written on a semi-logarithmic scale. 
Here, $\alpha=0.369$, $\phi (b)-\phi_B=-0.101$, 
and $v_0n_1(b)=2.73 \times 10^{-3}$ 
for  (a), while $\alpha=0.621$, 
$\phi (b)-\phi_B=0.214  $, 
$v_0n_1(b)=4.19 \times 10^{-2}$  for  (b). 
The lower panels of Fig.4 display the 
two functions $F_1(u)$ and $F_1^{\rm PB}(u)$ 
in Eq.(3.29) for the two cases.  
The excess  charge density in Eq.(3.28) is 
negative as 
 $\Delta\sigma=-0.02/\ell _{Bb}=
-0.03/\ell _{B0}$ for   (a),  
but is positive as  
$\Delta\sigma=0.33/\ell _{Bb}=
0.56/\ell _{B0}$ for  (b).

We then check the criterions (3.31) and (3.33). 
The left hand side of Eq.(3.31) 
is 1.73, while  that  of Eq.(3.33) 
is $-0.609$ for  (a) and  $0.755$ for (b). 
However, $F_1(b)=0.109$ for  (a) and $1.54$ for (b), so the 
 criterion (3.31) is not  
justified for (a) (see the discussion above Eq.(3.22)).

\subsubsection{First order  phase transition of ionization 
without salt}
%8 
\begin{figure}[t]
\begin{center}
\includegraphics[scale=0.5]{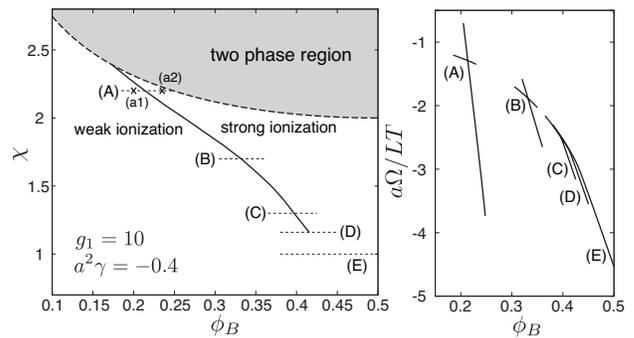}
\caption{Right: Phase diagram in the $\phi_B$-$\chi$ plane,  
where  $g_1=10$, $K_0v_0=8\times 10^{-6}$, 
 and $a^2\gamma= -0.4$  
for water-oil solvent without salt. 
Left: Normalized 
grand potential $a\Omega/ LT$  on 
 lines  (A)-(E). 
For (A)-(C) there are two branches of 
weak and strong ionization 
and   the  equilibrium  is given by the lower branch, while 
for (D) and (E) there is only one branch. 
}
\end{center}
\end{figure} 

% 10
\begin{figure}[htbp]
\begin{center}
\includegraphics[scale=0.59]{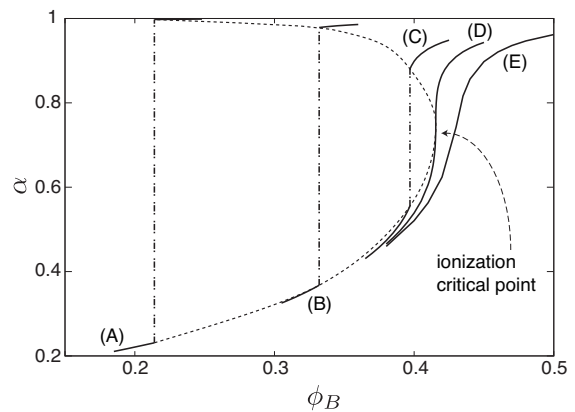}
\caption{Degree of ionization $\alpha$ vs $\phi_B$ 
 without salt   along the lines 
(A)$\sim $(E)  in Fig.5, where   
 $g_1=10$, $K_0v_0=8\times 10^{-6}$, 
 and $a^2\gamma= -0.4$. The discontinuity (broken line segments) 
vanishes at  the critical point of ionization.    
}
\end{center}
\end{figure}

With increasing $\phi_B$ we predict 
a first order phase  transition 
from weak to strong ionization. 
We suppose  that the backbone is hydrophobic, 
the ionization is weak in the pure second 
component, and the counterions 
are strongly hydrophilic. 
Hence, in Figs. 5-9, we set  
$a^2\gamma=-0.4\times 10^{-6}$, 
 $v_0K_0=8$, and $g_1=10$.

We vary $\phi_B$ at fixed $\chi$ on the 
lines (A)-(E) in the left  panel of Fig.5, where 
 $\chi=2.2$ (A), 1.7 (B), 1.3 (C), 1.16 (D), 
and 1  (E). For $\chi>\chi^{\rm cri}$,  
there are two branches of  weak and strong ionization 
around a first order transition line  expressed as 
\be 
\phi_B= 
\phi_B^{\rm tra}(\chi). 
\en  
It  starts  from a point given by 
$(\phi_B, \chi)=(0.1736, 2.39)$ 
on the solvent  coexistence curve and  ends  at 
a critical point given by 
$(\phi_B, \chi)=(\phi_B^{\rm cri}, \chi^{\rm cri})=
(0.415, 1.16)$, where $\phi_B^{\rm cri}=\phi_B^{\rm tra}
(\chi^{\rm cri})$. In the right  panel of Fig.5,   
the grand potential $\Omega$ in Eq.(3.13) is calculated, which  
is lower on the equilibrium  branch  
than on the metastable one. 
 In Fig. 6,  we show $\alpha$ vs $\phi_B$ 
  on  the lines (A)-(E), 
where  (A)-(C) pass through the 
transition line, 
(D) passes through the critical point, and (E) is 
 a  supercritical path. 
This ionization transition is 
analogous to the prewetting phase transition on a  
wall  \cite{Cahn,Ebner,Evans}, 
where a first order phase transition line  also 
starts from  the coexistence curve 
ending  at a critical point.

%11
\begin{figure}[htbp]
\begin{center}
\includegraphics[scale=0.42]{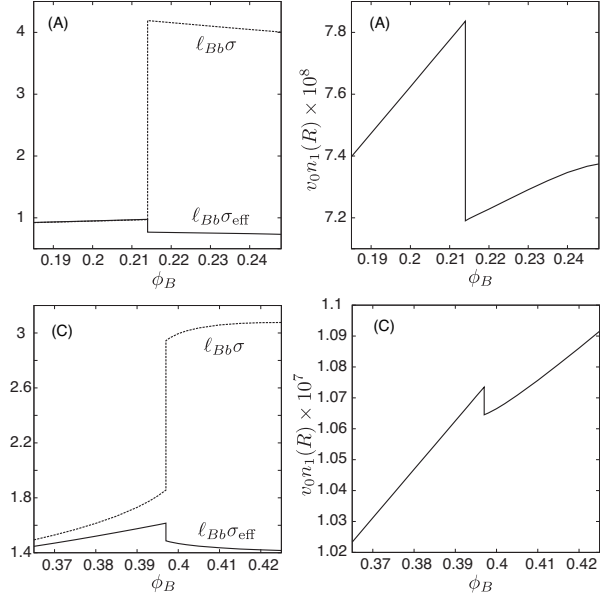}
\caption{Charge density $\sigma$ and effective  
charge density $\sigma_{\rm eff}$ 
in Eq.(3.19)  on the rod around the transition 
in units of $\ell_{Bb}^{-1}$ (left). 
Counterion density 
$n_1(R)= \Pi/T$ on the outer surface 
in units of $v_0^{-1}$ (right).  
 These are   plotted as   functions of $\phi_B$ on the line  
(A) (upper plates) and on the line 
(C) (lower plates) in Fig.5. 
In the weakly ionized phase,   
 $\ell_{Bb}\sigma$ 
is smaller than unity for (A) 
and larger than unity for (C), 
leading to larger discontinuities for (A) 
than for (C). 
}
\end{center}
\end{figure}

%12
\begin{figure}[htbp]
\begin{center}
\includegraphics[scale=0.42]{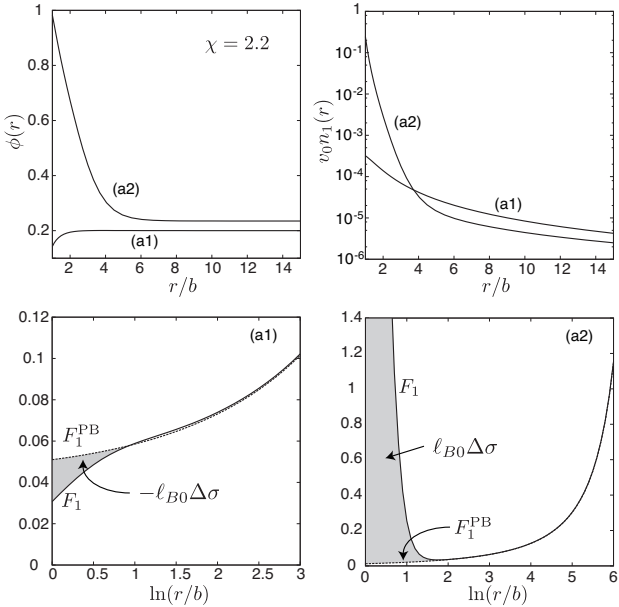}
\caption{Profiles without salt  at the two points 
(a1) (weakly ionized state)  and (a2) (strongly ionized state) 
 on the line (A)  in Fig.5. Top plates:
 $\phi(r)$ (left) and  
$v_0n_1(r)$ (right) vs $r/b$. 
Bottom plates: $F_1$ and $F_1^{\rm{PB}}$ vs $\ln(r/b)$ 
for (a1)(left) and  (a2)(right). 
From Eq.(3.28)  the area of the gray region 
is equal to $-\ell _{B0}\Delta \sigma$ for (a1) 
and to  $\ell _{B0}\Delta \sigma$ for (a2). 
}
\end{center}
\end{figure}

In the left panels of 
Fig.7, we   show the charge densities   
$\sigma$ and $\sigma_{\rm eff}$  multiplied by 
$\ell_{Bb}$  on the lines (A) and (C), 
where $\ell_{Bb}$ is defined in  Eq.(3.22) 
and $\sigma_{\rm eff}$ by Eq.(3.27). 
In these cases,  $\ell_{Bb}\sigma$ increases 
but $\ell_{Bb}\sigma_{\rm eff}$ decreases 
at the transition from weak to strong ionization, 
The counterions are more strongly attracted 
to the rod in the strongly  ionized state
than in the weakly  ionized state (see  Fig.8 below). 
The  right panels of Fig.7 display   the counterion density 
  $n_1(R)=\Pi/T$   on the outer surface  as a 
function of $\phi_B$.  Counter-intuitively,  
$\Pi$ decreases discontinuously at the transition 
with increasing $\alpha$. 
The jump of $\Pi$ at the transition is of order 
$8\%$ on the line (A) and $1\%$ on the line (C) 
in accord with  the modified Manning limiting law (3.30). 
Here   the effective Manning parameter 
$\ell _{Bb}\sigma _{\mathrm{eff}}$ is 
smaller  than unity on the line (A) and 
larger than unity on the line (C).

In Fig.8, we show the spatial  profiles   of $\phi(r)$, 
$n_1(r)$, $F_1(r)$,  and $F_1^{\rm{PB}}(r)$  
 at the two points (a1)  and (a2)  
 on the line (A) in Fig.5. 
The  left hand sides 
 of Eqs.(3.31) and (3.33) are equal to $5.83$  and $-0.877$ for 
(a1)  and to $6.13$ and $2.53$ for (a2), respectively. 
Here,  $\alpha=0.221$,  $\phi_b=0.143$, 
$v_0n_1(b)=3.20\times 10^{-4}$, and  
$\Gamma=0.706a^2$ at  (a1), 
while  $\alpha=0.998$,  $\phi_b=0.982$, $v_0n_1(b)=0.245$,  and 
$\Gamma=3.99a^2$ at   (a2). 
The upper left panel of Fig.8 
shows that the  water component is considerably 
depleted from the rod  at  (a1)  
but it covers  the rod almost completely  at  (a2). 
The upper right panel of Fig.8 
indicates that the counterions are 
more accumulated around the rod at (a2) than at (a1) 
such  that they are more depleted 
far from the rod at  (a2) than at (a1).
In the lower panels of Fig.8 the excess charge density 
$\Delta\sigma$ in Eq.(3.28) is $-0.003/\ell_{Bb}$
at (a1) and $3.32/\ell_{Bb}$ at (a2). 

\begin{figure}[htbp]
\begin{center}
\includegraphics[scale=0.46]{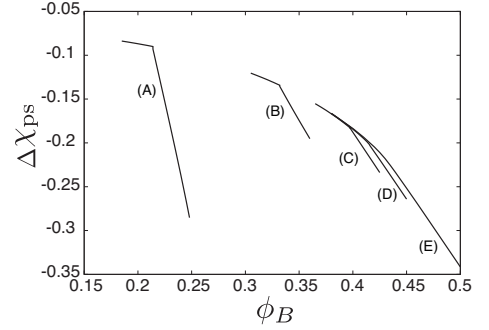}
\caption{Change of the effective 
polymer-solvent interaction parameter $\Delta\chi_{ps}$ 
due to ionization for a two component 
solvent without salt  along the paths (A)-(E) in Fig.5. 
}
\end{center}
\end{figure}

In Fig.9, we plot 
the change of the effective 
polymer-solvent interaction parameter 
$\Delta\chi_{ps}$ in Eq.(2.31)   along the paths (A)-(E). 
Its negativity  indicates  that the 
solvent quality becomes effectively   better 
with ionization and adsorption.  
In the brackets of Eq.(2.31),  
 the second term $\pi R^2n(R)$ 
is at most 10$\%$ of the first term $\Omega/TL$ 
and its discontinuity at the transition 
 is  very small.

We  comment on  the grand potential 
$\Omega$ in Eq.(3.13).  
At  the  transition,   
the composition part (the first line) 
increases  due to the layer formation, 
while  the  dissociation part 
(the first two terms in the second line) 
 decreases. The change of the electrostatic part   
is much smaller than these changes   
in the present case. 
At the transition point on the line A,  
these three parts (multiplied by $a$) 
are given by $(0.848, -0.775,-1.36)$  and 
$(11.0,-11.0, -1.28)$ in the weakly and strongly 
ionized states, respectively. 
%At the transition point on the line C, 
%the corresponding three parts  are  given by 
%$(2.56, -2.14, -2.93)$  and 
%$(5.17,  -4.72, -2.96)$, respectively. 
%% 13

%9
\begin{figure}[htbp]
\begin{center}
\includegraphics[scale=0.6]{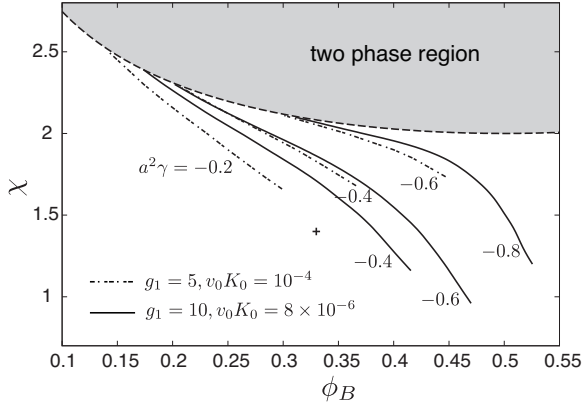}
\caption{ First-order  phase transition lines of ionization 
in the $\phi_B$-$\chi$ plane for various 
 parameter values without salt. 
Solid  lines:
$a^2\gamma= -0.4, -0.6,$ 
and $-0.8$ with  $g_1=7$ and $K_0v_0=8\times 10^{-6}$. 
Broken  lines: $a^2\gamma= -0.2, -0.4,$ 
and $-0.6$ with    $g_1=5$ and $K_0v_0= 10^{-4}$.
Each line starts from 
a point on the solvent coexistence curve 
and ends at a critical point with decreasing 
$\chi$ in  the bulk one-phase region. 
}
\end{center}
\end{figure}

The  first order phase transition 
can occur   over a wide range of the parameters both 
without  and with  salt. 
In  Fig.10, the first order phase transition lines are displayed 
for various $g_1$ and $\gamma$ without salt, which 
start  from the solvent  coexistence curve to  end 
at  an  ionization critical point. These lines  are markedly 
 enlarged with increasing the hydrophilic 
solvation strength $(g_1 >0)$ 
and/or the rod  hydrophobicity $(\gamma <0)$. 
The transition with salt will be examined in future.

\subsubsection{Profiles near the solvent 
 coexistence curve without salt}

In the following we 
 examine  the profiles of $\phi(r)$ and $n_1(r)$ 
close to the water-poor branch of  
 the solvent coexistence curve  ($\phi_B \le 1/2$). 
We vary $\phi_B$ and $\chi$ 
 fixing  the other parameters  as in 
Figs.5-9. Namely,   
 $g_1=10$, $a^2\gamma=-0.4$, and $v_0K_0=8\times 10^{-6}$.

Figure 11 presents  the profiles  at two points,  
$(\phi _B,\chi)=(0.1730,2.392)$ and  $(0.1740,2.389)$,  
between the transition point $(\phi _B,\chi)=(0.1736, 2.390)$ 
on the coexistence curve. 
We recognize marked jumps at the transition. 
 That is, at the weakly ionized state 
at $\phi_B=0.173$,  we have 
$\alpha=0.202$, $\sigma=1.21 \ell _{B0}^{-1} =0.900 \ell _{Bb}^{-1}$, 
and $\Gamma=0.565 a^2$, 
while at the strongly ionized state 
at $\phi_B=0.174$,  we have 
$\alpha=0.999$, $\sigma=5.99 \ell _{B0}^{-1} =4.44 \ell _{Bb}^{-1}$, 
and $\Gamma=27.2 a^2$. 
The first component remains  repelled around  
the rod in the weakly ionized phase 
but is much attracted  around it 
in the strongly ionized phase.

%% 14
\begin{figure}[htbp]
\begin{center}
\includegraphics[scale=0.4]{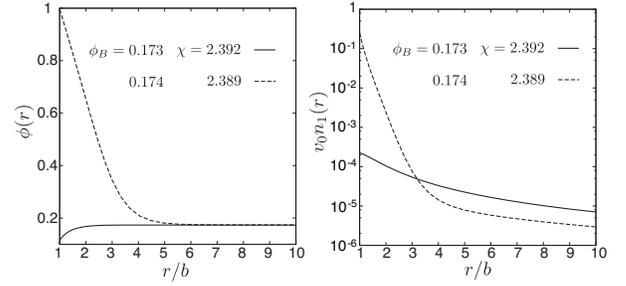}
\caption{Composition $\phi(r)$ and 
normalized counterion density $v_0n_1(r)$ 
on the  coexistence curve 
for $(\phi _B,\chi)=(0.1730,2.392)$ 
in the weakly ionized phase (bold lines) 
and   $(0.1740,2.389)$ in the strongly
 ionized phase (dotted lines). 
The line of the first order ionization 
phase transition starts 
 between these two points 
as in the right panel of Fig.5. 
}
\end{center}
\end{figure}
%% 15
\begin{figure}[htbp]
\begin{center}
\includegraphics[scale=0.4]{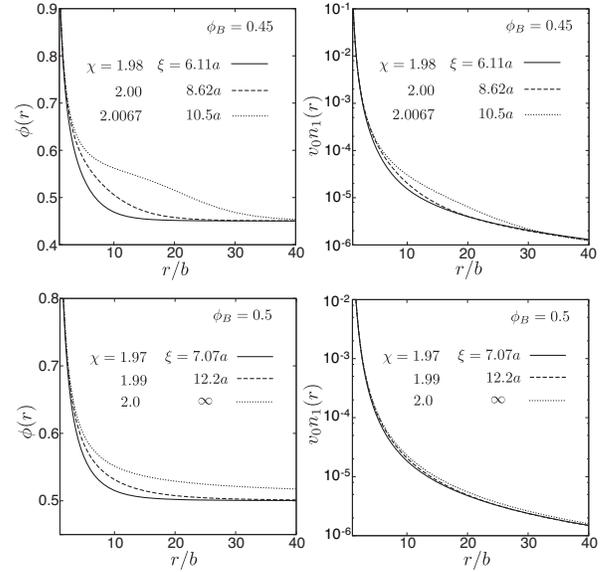}
\caption{Composition $\phi(r)$ and 
normalized counterion density $v_0n_1(r)$ 
near the solvent  coexistence curve 
for $0.45$ (top), and $0.5$ (bottom). 
Values of $\chi$ and $\xi$ are given within each panel. 
}
\end{center}
\end{figure}

In the upper panels of Fig.12, 
we set $\phi _B=0.45$  
 to obtain  
 $\Gamma/a^2=86.6$, $178$, and  $497$ 
for $\chi=1.9, 2.0$, and $2.0067$, respectively. 
The point $(\phi _B,\chi)=
(0.45,2.0067)$ is on the coexistence curve.  
In the lower panels, we set $\phi_B=0.5$ 
to obtain $\Gamma/a^2=88.9, 211.0,$ and $8691$ 
for $\chi=1.97, 1.990$, and $2.0$, respectively. 
A shoulder in $\phi(r)$ at  $(\phi_B,\chi)=(0.45,2.03)$ 
(left in the top)  represents  a water-rich layer 
varying gradually. 
At the  solvent critical point,  
the excess deviation  $\delta\phi(r)= 
\phi(r) -\phi_B$ decays as  
$\delta\phi (r) \sim r^{-1}$ 
for $r/a \ls 10$ and $\delta \phi (r) \sim r^{-2}$ 
for $10\ls r/a \ls 50$.  
In all these cases, 
the dissociation is nearly complete or $\alpha\cong 1$.

Marcus {\it et al.} \cite{Tsori1}   
obtained  a water-rich  layer 
separated from  the  water-poor bulk region  
by a sharp boundary 
around  a rod or a sphere. 
In our scheme,  a  layer with an interface  follows 
  in  the low density limit of  
the counterions (realized for very large $R$)
and under the conditions   
$\gamma=\Delta_1=0$ (which yields the boundary 
condition  $\phi'(b)=0$ from Eq.(3.17)).   
In addition, if the bulk region 
is  metastable (inside the 
solvent  coexistence curve), 
a  layer around a  chain 
can trigger phase separation.  This 
is analogous to the nucleation process 
around a charged  particle in metastable polar fluids 
\cite{Kitamura}.

\section{Summary and Remarks}

In this paper,   the ion distributions  have been   examined 
around a charged rod 
%in the mean field theory 
when the  degree of ionization $\alpha$ 
 is  a fluctuating quantity.  
In mixture solvents, the effect of the preferential solvation  
has been investigated.

In Sec.II,  a long ionizable 
rod in a one component solvent has been treated,
% without and with salt
 where  the dissociation process 
 gives rise to the free energy contribution 
$F_d$ in Eq.(2.8). Minimization of 
the grand potential $\Omega$ in Eq.(2.9) 
has then yielded  $\alpha$ obeying the mass action law 
and the  charge distribution  $n_1(r)$ obeying 
the Poisson-Boltzmann equation. 
In Subsec.IIB, we have examined the counterion density 
at the rod surface  without salt 
on the basis of the 
exact results  \cite{Fuoss}. 
For considerably  large $M=\ln(R/b)$,  
 $\alpha$ is determined by  Eq.(2.25) 
for $(1-\sigma\ell_B)M\gg 1$
and by  Eq.(2.26) for $(\sigma\ell_B-1)M\gg 1$. 
All the quantities sensitively depend 
on the parameter $A (\propto K_0)$ in Eq.(2.22) 
as in Fig.2.  In the limit $M \gg 1$, 
 $\alpha$ tends to unity in the former and 
to a well-defined limit in the range 
$(\ell_B\sigma_0)^{-1}<\alpha<1$ in the latter as in Fig.3.

  In Subsec.IIIA, we have 
generalized our theory in Sec.II to describe 
the ionization equilibrium  in mixture  solvents. 
The additional free energy 
 is $\Delta F$ in Eq.(3.1) for the composition $\phi$, 
which includes  the solvation couplings  with ions ($\propto g_i$) 
and the ionized monomers ($\propto \Delta_1$). The Manning limiting law for the osmotic pressure (A4)
 is modified to Eq. (3.30).
Though still fragmentary, 
Subsec.IIIB has 
 presented numerical results 
without salt, where  the solvent 
consists of a water-like component and a less polar component. 
The  counterions and the 
charged monomers are  hydrophilic  with 
 $g _1>0$ and $\Delta _1 >0$.  
Even if a rod is hydrophobic with $\gamma<0$, 
it becomes effectively hydrophilic with ionization as  in Fig.4. 
We have found 
 a first order phase transition of ionization 
 for hydrophilic counterions  
around a hydrophobic rod as in Figs.5-8. 
At the  transition from weakly to strongly 
ionized states, the number of the counterions increases,  
 but the osmotic pressure has decreased   
  in our examples as  in Fig.7.  
The polymer-solvent interaction 
parameter $\chi_{ps} $   decreases upon 
ionization as  in Fig.9. We have examined the composition and 
counterion profiles  at the crosspoint of the 
ionization transition line and  the coexistence curve 
 in Fig.11 
and near the coexistence curve 
in the strongly ionized phase in Fig.12. 
The  adsorption $\Gamma$ 
of the composition in Eq.(3.34) much 
increases  near the coexistence curve 
in the strongly ionized phase. 
The adsorption is long-ranged 
near the solvent criticality.  
In Appendix B, we will  add  a salt, where 
 $\alpha$  is a decreasing function of  
  the salt density. 
In Appendix C, we will  examine 
  the solvation shell formation 
 at  small content  of a  polar  solute.

We make  further  remarks on  the first order phase 
transition of ionization of an ionizable  rod. 
Here we suppose that a polymer chain can be in an expanded 
 coil state in a weakly ionized state 
and  in a more expanded state 
after the transition  without coil-globule transition. 
We have calculated inhomogeneities 
perpendicular to the chain, 
but   those   along the 
chain  should also be important 
and can well 
alter  the nature of the transition.

Finally, we 
 mention   previous and proposed 
experiments.
(i) Many authors have studied  the conformation of  a neutral polymer  
near the solvent critical point \cite{Brochard,mixture,Sumi}.  
The effect of the critical fluctuations 
 is much  more 
enhanced on a charged polymer  
in  a polar binary mixture. 
(ii) In near-critical 
binary mixtures with salt \cite{Anisimov} 
and in  polyelectrolytes \cite{Ise,Kuil,Amis}, 
 ion-induced aggregates 
have been observed, where  
relevance should be  the preferential solvation.   
(iii) We may replace  hydrophilic counterions 
by hydrophobic ions \cite{Osakai,Hung} such as  
tetrarphenylborate BPh$_4^{-}$.  
For hydrophilic  and hydrophobic  ion pairs, 
 the preferential solvation can be 
 much  stronger  than for hydrophilic ion pairs \cite{OnukiJCP}. 
 Sadakane {\it et al.} 
 \cite{Sadakane} added   NaBPh$_\mathrm{4}$ 
into  a binary mixture to obtain   mesophases.  
(iv) Ionic surfactants can be adsorbed to DNA 
 even if their bulk concentration 
is very small \cite{Yoshi1,Shirahama}. 
The resultant complex 
can be solubilized in organic solvents 
such as ethanol \cite{cationic}. Kuhn {\it et al.} predicted that 
this  adsorption can occur as a first order phase transition 
\cite{Levin1}.

In future work, we will  treat two parallel rods 
in mixture solvent,  
between which there can arise 
attraction mediated by the composition 
fluctuations.

\acknowledgments
{ This work was supported by Grants-in-Aid 
for scientific research 
on Priority Area ``Soft Matter Physics" 
and  the Global COE program 
``The Next Generation of Physics, Spun from Universality and Emergence" 
of Kyoto University 
 from the Ministry of Education, 
Culture, Sports, Science and Technology of Japan. 
The authors thank K. Yoshikawa, K. Nishida, 
T. Sumi, Y. Masubuchi, 
and  Y. Yamasaki for valuable discussions.    
}

\vspace{2mm} 
{\bf Appendix A: Calculations  
for one-component solvent without salt
}\\
\setcounter{equation}{0}
\renewcommand{\theequation}{A\arabic{equation}}

We show exact results for one-component solvent  without salt 
\cite{Fuoss,An}.
The function $F_1(r,\sigma)$ in Eq.(2.16) 
depends on $\hat{r}\equiv r/R$ 
logarithmically. Depending on whether 
$\sigma< \sigma^* $ or $\sigma> \sigma^* $ it behaves  as 
\bea
&&\hspace{-0.85cm}
\frac{F_1(r,\sigma)}{B^2} =  
[ \sinh (B\ln \hat{r}- \tanh ^{-1}B)]^{-2} 
\quad (\sigma< \sigma^* )\nonumber\\ 
&&=   [ \sin (B\ln \hat{r}- \tan ^{-1}B)]^{-2}\quad (\sigma> \sigma^* ).
\ena
The parameter $B$ is the  solution of   the   equation, 
\bea
&&\hspace{-0.75cm}
\frac{1-\ell_B\sigma}{B} = 
  \coth [BM+ \tanh ^{-1}B] \quad (\sigma < \sigma^*)\nonumber\\ 
&&\hspace{0.45cm}  =  \cot [BM+ \tan^{-1}B] \quad (\sigma > \sigma^*) ,
\ena
where we may set $B\ge 0$.
At $\sigma = \sigma^*$, we have $B=0$ and $F_1(r,\sigma^*)=
(1-\ln \hat{r})^{-2}$ 
for any $M>0$.  At $r=b$ it holds 
\be 
F_1(b,\sigma)=(\ell_B\sigma-1)^2\mp B^2,  
\en 
where $-$ is for $\sigma <\sigma ^*$ 
and  $+$ is for $\sigma >\sigma ^*$ 
in the right hand side. 
For $M\gg 1$ 
we find 
$B\cong 1-\ell_B\sigma$ for $-q \gg 1$ and 
$B\cong \pi/M$ for $q \gg 1$ 
in terms of $q$ in Eq.(2.20).

At  $r=R$  we obtain   
the large $M$ behavior  
$F_1(R,\sigma) \cong 1- (1-\ell_B\sigma)^2$ 
for $\ell_B\sigma<1$ and 
$F_1(R,\sigma) \cong 1$ for $\ell_B\sigma>1$.   
We may rewrite  this result 
 in terms of the osmotic pressure 
$\Pi$ in Eq.(2.24).  
In the limit $M\to \infty$, it follows 
the Manning limiting law for the osmotic pressure,  
\bea 
{\Pi} &\cong& Tn_p(1-\ell_B\sigma/2) \quad (\ell_B\sigma<1)
\nonumber\\
&\cong&  Tn_p/2\ell_B\sigma  \qquad (\ell_B\sigma>1),
\ena
where $n_p \equiv  
 \sigma/\pi R^2$.   
For  $\ell_B \sigma>1 $,  
 $\Pi$ saturates at $T/2\pi\ell_BR^2$.  
Thus a fraction of $1-(\ell_B\sigma)^{-1}$ 
of the  counterions are apparently 
localized around  the rod 
(the Manning-Oosawa counterion condensation) 
\cite{Manning,Oosawa,Barrat,Levin,Volk,Holm,Rubinstein,Baigl1,An,Netz}.

We calculate $\Omega$ in Eq.(2.13). 
Since  the electric field  is given by 
$E
%=- \frac{T}{e} \frac{dU}{dr} 
={T}{e}^{-1} 
[{d}(\ln F_1)/dr -{2}/{r}], 
$ the electrostatic energy  
 becomes 
\be 
F_e = TL\sigma 
+ TL\ell_B^{-1} {\cal F}_e (\ell_B\sigma),
\en 
where 
${\cal F}_e(s)$ with $s=\ell_B\sigma$ appears in Eq.(2.27). 
From  Eqs.(A1) and (A2) some calculations  
yield\cite{Netz}   
\bea
{\cal  F}_e(s)  
 &=&
(1+B^2)M+\ln \bigg[1+\frac{s^2-2s}{1-B^2}\bigg] \quad  (\sigma<\sigma^*)
 \nonumber\\
&&\hspace{-15mm} =
(1-B^2)M+\ln \bigg[1+\frac{s^2-2s}{1+B^2}\bigg]\quad  (\sigma>\sigma^*) .
 \ena 
For large $M$ we  obtain 
approximate expressions,  
% \bea
% {\cal F}_e(s) 
% &\cong &s^2M+2 \ln \bigg[\frac{2-2s}{2-s}+{M}^{-1}\bigg] 
% \quad (q<-1) \nonumber \\
% &\cong&   M+ 2\ln 
% (s-1 + M^{-1}) \quad (q >-1), 
% \ena
\bea
{\cal F}_e(s) 
&\cong &s^2M+ \ln \bigg[\frac{2(1-s)^2}{2-s}+{M}^{-1}\bigg] 
\quad (q<-1) \nonumber \\
&\cong&   M+ 2\ln 
(s-1 + M^{-1}) \quad (q >-1), 
\ena
where $q=M(s-1)$. In deriving the first line we have used 
Eqs.(2.21) and (A3). 
We introduce  $M^{-1}$ on the right hand sides  
to avoid the logarithmic divergence at $s=1$. 
At $\sigma=\sigma^*$ we have 
%$s=M/(1+M)$ and 
${\cal F}_e= M -2\ln(1+M)$.

\vspace{2mm} 
{\bf Appendix B: Charged rod in one-component solvent with salt}\\
\setcounter{equation}{0}
\renewcommand{\theequation}{B\arabic{equation}}

Here we  
examine the counterion density 
and the degree of ionization in 
one-component solvent with added salt, 
which is completely dissociated into cations and anions.
We assume that the cations from the salt are 
of the same species as the counterions  from 
 the rod. For example,  we suppose the 
chemical reactions: 
\bea
&&-\mathrm{COOX} \rightleftharpoons -\mathrm{COO}^- + \mathrm{X}^+  
\quad (\mathrm{rod~surface})\nonumber\\
&&\mathrm{XCl} \to \mathrm{X}^+ +\mathrm{Cl}^- \quad 
(\mathrm{salt~in~bulk}) \nonumber,
\ena
where X=H or Na. 
The cation and anion   densities 
are denoted by $n_1$ and $n_2$, respectively. 
We treat  the  anion density at $r=R$  
as a control parameter and  write it  as $n_{B}$. 
In this case, the electric field created by 
the positive charges  on the rod is screened 
for $r>k_D^{-1}$, where 
\be 
k_D= (8\pi \ell_B n_B)^{1/2}
\en 
is the 
Debye wave number far from the rod.  For $r\gg k_D^{-1}$, 
the system is homogeneous with $n_1(r)\cong n_2(r)$.  
If   $R\gg k_D^{-1}$, we 
obtain the results in the limit $R\to \infty$. 
By  setting  $\Phi(\infty)=0$, 
we may write the ion densities as 
 $n_1=n_Be^{-e\Phi/T}$ and 
$n_2=n_Be^{e\Phi/T}$, where  $n_1(\infty)=n_2(\infty)=n_B$.
We rewrite the Poisson-Boltzmann equation as  
\be
 \bigg (\frac{d^2}{dr^2}+ 
\frac{1}{r}\frac{d}{dr}\bigg)  \frac{e\Phi}{T} = 
k_D^2 \sinh\bigg(\frac{e\Phi}{T}\bigg),
\en 
 The boundary condition is 
given by Eq.(2.6) at $r=b$.

% 5
\begin{figure}[t]
\begin{center}
\includegraphics[scale=0.42]{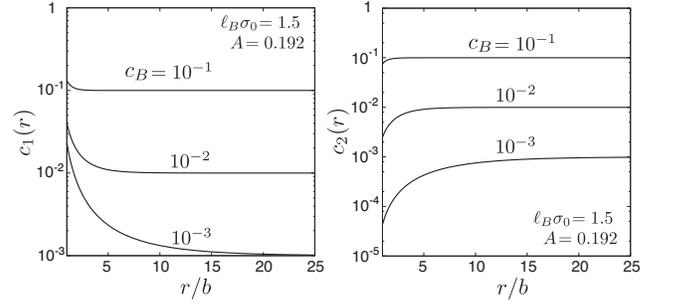}
\caption{Normalized density of cations $c_1(r)=
 \ell _B b^2 n_1(r)$ (left) and anions $c_2(r) = 
\ell _B b^2 n_2(r)$ (right) vs $r/b$ with salt at  
 $\ell _B \sigma _0=1.5$ and $A=0.192$.  
The salt density is given by    
 $c_B= \ell _B b^2 n_B =
10^{-1}, 10^{-2}$ and $10^{-3}$ .}
\end{center}
\end{figure}
% 6
\begin{figure}[t]
\begin{center}
\includegraphics[scale=0.47]{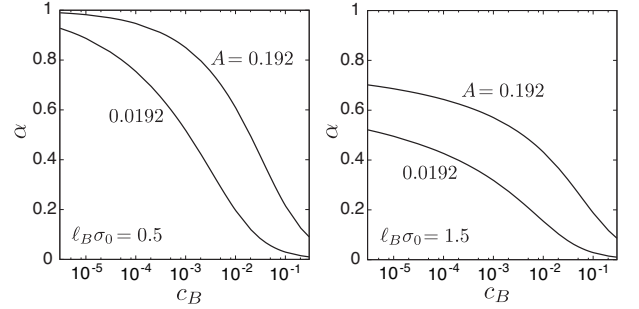}
\caption{Degree of ionization 
$\alpha$ as a functions of salt concentration 
$c_B \ell _B b^2 n_B$  with salt 
on a semi-logarithmic scale for 
$A=0.192$ and $0.0192$.  
Here  $\ell_B\sigma_0=0.5<1.0$ (left) 
and  $\ell_B\sigma_0=1.5>1.0$ (right). 
}
\end{center}
\end{figure}

We numerically solved the above 
Poisson-Boltzmann equation  
and the ionization equation Eq.(2.14) to obtain 
equilibrium $n_1(r)$, $n_2(r)$, and $\alpha$.  
In Fig. 13,  we plot the normalized ion densities, 
\be
c_1(r) = \ell _Bb^2 n_1(r), \quad  
c_2(r) =  \ell _Bb^2 n_2(r),  
\en 
which tend to $c_B\equiv  \ell _Bb^2 n_B$ 
for $r\gg k_D^{-1}$. 
We set 
$\ell_B\sigma _0=1.5$ and $A=0.192$. 
Here $c_1(r)$ increases 
and $c_2(r)$  decreases  near the rod.  
In Fig. 14, 
 we show $\alpha$ as a function of $c_B$ for 
$\ell _B\sigma _0=0.5$ and $1.5$. 
We find that $\alpha$ decreases with 
increasing the salt concentration $c_B$. 
%This is consistent with Fig. 13, where  $n_1(b)$  becomes  
%larger with increasing $c_B$. 
We also find $\alpha \to 1$ 
as $c_B\to 0$ for $\ell _B\sigma _0 =0.5<1$, 
while $\alpha$ does not approach unity as 
$c_B\to 0$ for $\ell _B\sigma _0=1.5>1$. 
This is consistent with the behavior of 
$\alpha$ in the limit $M\to \infty$ 
in the salt-free case discussed 
in Subsec.IIB. 
In the limit $c_B \to 0$ (with $R=\infty$),  
$\alpha$  approaches 
 unity for $\ell_B\sigma_0<1.0$ (though 
this limiting  behavior is not seen for  $A=0.0192$ 
in the left panel), while it 
 tends to a constant 
in the range  $(\ell_B\sigma_0)^{-1}<\alpha<1$ 
 for $\ell_B\sigma_0>1.0$ 
as can be seen in the right panel.

Thus, with increasing the salt density, 
$\alpha$ is reduced  and the electrostatic interaction is 
screened. Experimentally, with addition of salt,  
 highly expanded polyelectrolyte
coils  have been observed to 
 shrink, eventually 
resulting in  precipitation of 
the chains (phenomenon known as "salting out" of 
polyelectrolytes \cite{Volk}).

\vspace{2mm} 
{\bf Appendix C: 
Water adsorption 
around hydrophilic ions 
at small water content 
}\\
\setcounter{equation}{0}
\renewcommand{\theequation}{C\arabic{equation}}

We here present a statistical theory of  
water adsorption to hydrophilic ions\cite{electro}.  
The system is 
in the cylindrical cell  $b<r<R$ and $0<z<L$. Each solvation shell  
consists   of $\nu$ water molecules, where 
$ \nu=1,\cdots, S$ with $S$ 
being the maximum number. The 
 binding energy is $Tw_\nu$. 
If  $w_\nu \gg 1$, the adsorption  
can be significant  even for small 
bulk water composition $\phi_B$ (see Eq.(C8)).

The  number of ionized monomers  is  
$N_I= L\sigma$.  The numbers of $\nu$-clusters  
composed of $\nu$ water molecules 
are   $\beta_\nu N_I$.   
The total number of the hydrated   ionized monomers 
is then $\beta N_I$ with 
\be 
\beta=\sum_{\nu} \beta_\nu<1.
\en  
The fractions $\beta_\nu$ are determined by 
minimization of  the free energy of the form, 
\bea 
\frac{F_s}{T}&=& {N_0} 
 \phi_B(\ln\phi_B-1)+N_I (1-\beta)\ln (1-\beta) \nonumber\\
&&+  N_I \sum_{\nu} 
 \beta_\nu(\ln \beta_\nu-w_\nu)  ,
\ena  
where $N_0=V/v_0$ with  $V= \pi (R^2-b^2) L$ being  the 
volume occupied by the solvent.   The total number of the 
water molecules is fixed as  
\be 
N_0 \phi_B + N_I\sum_{\nu} 
 \nu \beta_\nu =N_0\phi_B^0,
\en  
where $\phi_B^0$ is the  volume fraction without 
adsorption. From $\partial F_s/\partial \beta_\nu=0$ under 
Eq.(C3) we obtain 
\bea 
{\beta_\nu}&=&({1-\beta})\phi_B^\nu e^{w_\nu}, \\ 
\beta&=&  1-1/[1+\sum_{\nu}\phi_B^\nu e^{w_\nu}].
\ena
For $N_I\ll N_0$ or for $R\gg b$, 
we may set  $\phi^0-\phi \ll \phi$  even if $\beta$ approaches unity.
We then calculate the excess free energy  
$\Delta F_s=F_s-T{N_0} \phi_B^0(\ln\phi_B^0-1)$ 
due to the water adsorption. 
Some calculations give 
\bea 
\frac{\Delta F_s}{T}&=& N_0\phi_B^0\ln(\phi_B/\phi_B^0) 
+ N_I[\phi_B^0-\phi_B + \ln(1-\beta)]\nonumber\\
&\cong&  -N_I\ln[1+\sum_{\nu}  \phi_B^\nu e^{w_\nu}]. 
\ena
Here we  set $\ln(\phi_B/\phi_B^0)\cong 
\phi_B/\phi_B^0-1$ in the first line to obtain 
 the second line for $N_I\ll N_0$.

The formation of solvation shells around 
the counterions may  be calculated  in the same manner. 
Let  $Tw_\nu'$  with $\nu=1,\cdots,S'$ 
be the binding energy of 
 $\nu$-clusters of  water molecules.  
For monovalent counterions,  we find the  
 free energy decrease 
in the same form as that in Eq.(C6) 
with $w_\nu$ being replaced by $w_\nu'$. 
For   a sufficiently small water 
density outside the solvation shells,   
we   may use the results of one-component solvents in Sec.II  
 if $\Delta_0$ is replaced  by  
\be 
{\tilde\Delta}_0 =\Delta_0-
\ln \bigg\{\bigg[1+\sum_{\nu=1}^S \phi_B^\nu   
 e^{w_\nu}\bigg] 
 \bigg[1+\sum_{\nu=1}^{S'}
\phi_B^\nu e^{w_\nu'}\bigg ]\bigg\}.
\en 
The dissociation constant $K_0$ in Eq.(2.16) 
is changed to  
$\tilde{K}_0 = n_{01}\exp({-\tilde{\Delta}_0})$. 
and the parameter  $A$  in   Eq.(2.23) is replaced by 
$\tilde{A}=2\pi\ell_B b^2 {\tilde K}_0$.

Let the maximum  of $w_\nu/\nu$ 
and  $w_\nu' /\nu$ be  $w_s$. 
Significant ionization enhancement occurs   for 
\be 
\phi_B\gg \exp({-w_s})   ,
\en 
where the right hand side is small for 
$w_s \gg 1$.


\begin{thebibliography}{99}
%\begin{thebibliography}{99}


\bibitem{Manning} G.S. Manning,  
  J. Chem. Phys. {\bf 51}, 924 (1969); 
J. Phys. Chem. B  {\bf 111}, 8554 (2007). 


\bibitem{Oosawa} 
F. Oosawa, Polyelectrolytes (Marcel Dekker, New York, 1971).



\bibitem{Barrat} J.L. Barrat and J.F. Joanny, 
 Adv. Chem. Phys. XCIV, I. Prigogine, S.A. Rice Eds.,
John Wiley $\&$ Sons, New York 1996.  


\bibitem{Levin} 
Y. Levin, Rep. Prog. Phys. {\bf 65}, (2002) 1577.

\bibitem{Volk} 
N. Volk, D. Vollmer,  M. Schmidt, 
W. Oppermann, and K. Huber, 
Adv. Polym.  Sci. {\bf 166}, 29 (2004).


\bibitem{Holm} 
C. Holm,  J. F. Joanny,   K. Kremer, 
 R. R. Netz,  P. Reineker,  C. Seidel, 
T. A. Vilgis, and R. G. Winkler, 
Adv. Polym.  Sci. {\bf 166}, 67 (2004).



\bibitem{Rubinstein} A.V. Dobrynin and M. Rubinstein, 
 Prog. Polym. Sci. {\bf 30}, 1049 (2005).  


\bibitem{Baigl1}
W. Essafi, F. Lafuma, D. Baigl, and C. E. Williams, 
Europhys. Lett. {\bf 71},  938 (2005).


\bibitem{An} D. Andelman, Proceeding of the Nato ASI  and SUSSP on "soft condensed matter in molecular and cell biology", (2005), ed. by W. Poon and D. Andelman, (Taylor \& Francis, New York, 2006), pp. 97-122.

\bibitem{Netz} A.  Naji  and R. R. Netz, 
Phys. Rev. E {\bf 73}, 056105 (2006).


\bibitem{Me} R.  Messina, 
J. Phys.: Condens. Matter {\bf 21},  113102 (2009). 

\bibitem{Joanny} E. Raphael and 
 J. F. Joanny, Europhys. Lett.  {\bf 13}, 623 (1990).


\bibitem{Bu1} I. Borukhov, D. Andelman, and 
H. Orland, Europhys. Lett.{\bf 32}, 499 (1995). 



\bibitem{Bu2}  I. Borukhov, D. Andelman, R. Borrega, M. Cloitre, 
L. Leibler,  and 
H. Orland, J. Phys. Chem. B  
{\bf 104}, 11027 (2000). 

\bibitem{Muth} 
M. Muthukumar, J. Chem. Phys. {\bf 120},  9343 (2004). 
\bibitem{Burak} 
Y. Burak and R. R. Netz, J. Phys. Chem. B {\bf 108},  4840 (2004). 
This paper shows that 
the electrostatic interaction 
among the charged monomers 
 gives rise to correlated ionization. 



\bibitem{Onuki-Okamoto} 
A. Onuki and R. Okamoto,  
J. Phys. Chem. B, {\bf 113}, 3988 (2009).

%\bibitem{Krama} E.Y. Kramarenko, A.R. Khokhlov, 
%and K. Yoshikawa, 
% Macromol Theory Simul. {\bf 9}, 249 (2000). 

%%%%%%%%%%%%%%%%%
\bibitem{Zimm} C. B. Post and B. H.  Zimm, 
 Biopolymers {\bf 21}, 2139 (1982). 
\bibitem{Bloomfield} 
P. G. Arscott, C. Ma, J. R. Wenner  and V. A. Bloomfield, 
Biopolymers, {\bf  36},  345 (1995). 
% Andelman cite
\bibitem{Rau} 
A. Hultgren and D. C. Rau, 
Biochemistry {\bf 43}, 8272 (2004). 

\bibitem{Rauy} C.  Stanley and D. C. Rauy, 
Biophy. J.  {\bf  91}, 912 (2006). 


%zwitterionic species
\bibitem{Flock} S. Flock, R. Labarbe, and C.  Houssier, 
Biophysical Journal {\bf  70}, 1456 (1996). 



\bibitem{Baigl} D. Baigl and  K. 
Yoshikawa, J.  Biophys. {\bf 88}, 3486 (2005). 



\bibitem{Yoshi1} 
S. M. Mel'nikov,  V. G. Sergeyev,  and K. Yoshikawa, 
J. Am. Chem. Soc. {\bf 117}, 2401 (1995). 


\bibitem{Andelman1} 
D. Ben-Yaakov, D.  Andelman, D. Harries, and R.  Podgornik, 
J. Phys. Chem. B {\bf 113}  6001  (2009). 




%%%%%%%%%%%%%%%%

\bibitem{Is} J. N. Israelachvili,  
{\it Intermolecular and Surface 
Forces} (Academic Press, London, 1991). 



\bibitem{Onuki-Kitamura} A. Onuki and H. Kitamura, 
  J. Chem. Phys. {\bf 121}, 3143 (2004).


\bibitem{OnukiPRE} A. Onuki, Phys. Rev. E {\bf 73}, 021506 (2006).
\bibitem{OnukiJCP} A. Onuki, 
J. Chem. Phys.  {\bf 128}, 224704 (2008). 



\bibitem{Tsori} 
%G. Marcus, S. Samin, and Y. Tsori, 
%J. Chem. Phys. {\bf 129}, 061101 (2008).
Y. Tsori and L. Leibler, 
 Proc. Natl. Acad. Sci. U.S.A. 
 {\bf 104}, 7348  (2007)

\bibitem{Roij} 
M. Bier, J. Zwanikken, and R. van Roij, Phys. Rev. Lett. {\bf 101},
046104 (2008); 
J. Zwanikken, J. de Graaf, M.  Bier, 
and R. van Roij, J. Phys.: Condens. Matter {\bf 20}, 494238 (2008).

\bibitem{OnukiEPL} A. Onuki, Europhys. Lett. 
 {\bf 82}, 58002 (2008).



%%%%%%%%%%%%%%
\bibitem{Cahn} J. W. Cahn, J. Chem. Phys. {\bf 66} 3667 (1977).
\bibitem{Ebner}
C. Ebner and W. F. Saam, Phys. Rev. Lett. 38, 1486 (1977).
\bibitem{Evans}
P. Tarazona and R. Evans, Mol. Phys. 48, 799 (1983).



\bibitem{Onukibook} A. Onuki, {\it Phase Transition Dynamics} 
(Cambridge University Press, Cambridge, 2002)



\bibitem{Fuoss} R. M. Fuoss, A. Katchalsky, 
and S. Lifson, Proc. Natl. Acad. Sci. U.S.A., {\bf 37}, 579 (1951).


\bibitem{v0} For ideal gases  the entropic free energy 
is  of the form  
$Tn_i[ \ln (n_iv_{0i})-1]$ with 
 $v_{0i}= \hbar^3 (2\pi/m_i T)^{3/2}$, where 
 $\hbar$ is  the Planck constant and $m_i$ is  the particle 
mass. This term is larger than 
that in Eq.(2.7) by $n_i\Delta\mu_i$, where  
$\Delta\mu_i= T\ln(v_{0i}/v_{0})$ is a constant 
shift of the chemcal potential of the $i$ th ions. 
 




\bibitem{Osakai} T. Osakai and K. Ebina, 
J. Phys. Chem. B {\bf 102}, 5691 (1998). 
In water-nitrobenzene(Nr) in two phase coexistence, 
they measured  the number of hydrating  water molecules 
around  ions in a Nr-rich region   
 at room temperatures. 
The water number per ion in the Nr-rich phase 
was estimated to be  4 for Na$^+$, 
6 for Li$^+$, and 15 for Ca$^{2+}$, while it was 
nearly zero for hydrophobic ions. 
Their  experiment demonstrates 
 strong binding of water molecules to   
hydrophilic ions even at
 small water contents. 






\bibitem{electro} 
A. Hamnett, C. H. Hamann, and W. Vielstich, 
{\it Electrochemistry} 
(Vch Verlagsgesellschaft Mbh, 1998). 



\bibitem{Debye-Kleboth} P. Debye and K. Kleboth, 
  J. Chem. Phys. {\bf 42}, 3155 (1965).


\bibitem{Ohtaki} H. Ohtaki, Bulletin of the Chemical 
Society of Japan, {\bf 42}, 1573 (1969).

\bibitem{Bonn}
D. Bonn, D. Ross, S. Hachem, S. Gridel, and J. Meunier,  
Europhys. Lett. {\bf  58}, 74 (2002). 



\bibitem{Born} M. Born, Z. Phys. {\bf 1}, 45 (1920). 






\bibitem{Hung}  
Le Quoc Hung, J. Electroanal. Chem. 
{\bf 115}, 159 (1980).

 



\bibitem{Tsori1} 
G. Marcus, S. Samin, and Y. Tsori, 
J. Chem. Phys. {\bf 129}, 061101 (2008).
These authors assumed   the homogeneity of 
$h=f_0'(\phi) -\ve_1 E^2/8\pi$, where 
the counerions are absent 
and the gradient term is 
 neglected. Compare their $h$ and  our $h$ in  Eq.(3.11).  
In their theory,  
$E^2 (\propto r^{-2}$ for rods) 
plays the role of an inhomogeneous ordering field. 



\bibitem{Kitamura}
H. Kitamura and A. Onuki, J. Chem. Phys. {\bf 123}, 124513 (2005).

\bibitem{Brochard} 
P.G. de Gennes, 
J. Phys. (Paris)  {\bf 37}, 59  (1976). 
Polymers can interact with 
two solvent components asymmetrically. 
There arises a pairwise attractive 
interaction among the monomers 
mediated by the 
critical fluctuations. 
Similar attractive interactions 
were derived among ions 
in mixture solvents by one of the present 
authors\cite{OnukiPRE}.  

 
 



\bibitem{mixture} 
A.  Dondos and Y.  Izumi, 
Makromol. Chem. {\bf 181}, 701 (1980); 
C. A. Grabowski and A. 
Mukhopadhyay, Phys. Rev. Lett. 
{\bf 98}, 207801 (2007).

\bibitem{Sumi} 
T. Sumi, K.  Kobayashi, and H.  Sekino, 
J. Chem. Phys. {\bf 127}, 164904 (2007).


\bibitem{Anisimov} 
A. F. Kostko, M. A. Anisimov, and J. V. Sengers, 
Phys. Rev. E {\bf 70}, 
026118 (2004). 

\bibitem{Ise} H.  Matsuoka, D.  Schwahn, and N.  Ise, 
Macromolecules {\bf 24}, 4227 (1991). 
\bibitem{Kuil} J. J. Tanahatoe and M. E. Kuil, 
J. Phys. Chem. B {\bf 101}, 5905 (1997).
\bibitem{Amis} B. D. Ermi and E.  J. Amis,
Macromolecules {\bf  31}, 7378 (1998).

\bibitem{Sadakane} K. Sadakane, H. Seto, H. Endo, and M. Shibayama, 
J. Phys. Soc. Jpn., {\bf 76}, 113602 (2007); 
{K. Sadakane}, A.  Onuki, K.  Nishida, 
S.  Koizumi, and H.  Seto, 
preprint(arXiv:0903.2303v2).   


\bibitem{Shirahama} K. Shirahama, K. Takashima, 
and N. Takisawa, Bulletin of the Chemical 
Society of Japan, {\bf 60}, 43 (1987). 
%A.V. Gorelov, E.D. Kudryashov, J.-C. Jacquier, D.
%McLoughlin, K.A. Dawson, Physica A {\bf 249}, 216 (1998). 



\bibitem{cationic} 
K. Tanaka and Y.  Okahata, 
J. Am. Chem. Soc.,  {\bf 118}, 10679 (1996). 

\bibitem{Levin1} 
P. S. Kuhn, Y.  Levin, M. C. Barbosa,
Chemical Physics Letters {\bf 298}, 51 (1998). 




\end{thebibliography}
\end{document}